\documentclass[times,review,10pt,3p]{elsarticle}

\usepackage{amssymb}
\usepackage{amsmath}
\usepackage{tabularx,float,multirow,lineno}

\usepackage{tikz}                

\usepackage{graphicx}     
\usepackage{adjustbox,ctable,makecell}
\usepackage{tcolorbox}
\usepackage{mdframed}
\usepackage{threeparttable}
\usepackage{enumitem}
\usepackage{longtable}
\usepackage{booktabs}

\usepackage{etoolbox}
\BeforeBeginEnvironment{tabularx}{\centering\footnotesize}
\BeforeBeginEnvironment{tabular}{\centering\footnotesize}
\newcolumntype{+}{>{\global\let\currentrowstyle\relax}}
\newcolumntype{^}{>{\currentrowstyle}}
\newcommand{\rowstyle}[1]{\gdef\currentrowstyle{#1}#1\ignorespaces}

\newcolumntype{Y}{>{\centering\arraybackslash}X} 

\journal{Information and Software Technology}

\begin{document}
\begin{frontmatter}

\title{A Systematic Mapping on Software Fairness: Focus, Trends and Industrial Context}

\author[ets]{Kessia Nepomuceno\corref{cor1}}
\address[ets]{École de Technologie Supérieure,\\ Département de Génie Logiciel et des Technologies de l'Information, \\ 1100 Rue Notre-Dame Ouest, Montréal QC H3C 1K3, Canadá.}
\ead{kessia.cavalcanti-nepomuceno.1@ens.etsmtl.ca}

\author[ets]{Fabio Petrillo}
\ead{fabio.petrillo@etsmtl.ca}

\cortext[cor1]{Corresponding author}

\begin{abstract}

\noindent \textbf{Context:} Fairness in systems has emerged as a critical concern in software engineering, garnering increasing attention as the field has advanced in recent years. While several guidelines have been proposed to address fairness, achieving a comprehensive understanding of research solutions for ensuring fairness in software systems remains challenging. 

\noindent \textbf{Objectives:} This paper presents a systematic literature mapping to explore and categorize current advancements in fairness solutions within software engineering, focusing on three key dimensions: research trends, research focus, and viability in industrial contexts. 

\noindent \textbf{Methods:} We develop a classification framework to organize research on software fairness from a fresh perspective, applying it to 95 selected studies and analyzing their potential for industrial adoption.

\noindent \textbf{Results:} Our findings reveal that software fairness research is expanding, yet it remains heavily focused on methods and algorithms. It primarily focuses on post-processing and group fairness, with less emphasis on early-stage interventions, individual fairness metrics, and understanding bias root causes. Additionally fairness research remains largely academic, with limited industry collaboration and low to medium Technology Readiness Level (TRL), indicating that industrial transferability remains distant.

\noindent \textbf{Conclusion:} Our results underscore the need to incorporate fairness considerations across all stages of the software development life-cycle and to foster greater collaboration between academia and industry. This analysis provides a comprehensive overview of the field, offering a foundation to guide future research and practical applications of fairness in software systems.


\end{abstract}




\end{frontmatter}


\section{Introduction}
\label{introduction}


Some organizations, including companies along with government sectors, have found fairness issues in their software applications \cite{brun2018software}. These cases have highlighted and enforced the critical need for fair and unbiased systems, in different areas and sectors like hiring, loan approvals, and public policy implementations. As a result, fairness in software has become a central concern in both academic research and industry practices.
This growing awareness has motivated investigation into the fairness dominion, leading researchers to explore its implications and applications in different areas.  


Throughout the literature, significant efforts have been made by the scientific community to define and understand fairness in the systems \cite{verma2018fairness, saleiro2018aequitas, baresi2023understanding}. We can say that fairness in our context refers to the goal of ensuring that machine learning models or software systems do not perpetuate, reinforce, or amplify existing societal biases \cite{schwartz2022towards, nepomuceno2025ai}. Achieving fairness is crucial in prompting ethical and responsible AI, especially as these systems are increseangly applied in high-stakes domains, where the impacts of bias can be deep.


The discussion of fairness in computing traces its roots back to early concerns about the unintended consequences of algorithmic decision-making. In the 1960s and 1970s, as automated systems began to influence sectors such as banking, employment, and criminal justice, questions about the fairness and transparency of these systems emerged \cite{boguslaw1968privacy, welfare1973records}. However, it was not until the late 20th and early 21st centuries, with the rapid growth of AI and machine learning, that fairness in software became a major focus of both academic and public debate \cite{barocas2016big, o2017weapons}.

One of the earliest high-profile cases illustrating the risks of unfair software was the 1980s U.S. credit scoring system, which disproportionately disadvantaged minority applicants \cite{ladd1998evidence}. Similar cases resurfaced in the 2000s, including fairness issues in hiring algorithms, and predictive policing systems that unfairly targeted minority communities \cite{andrews2022automating, ferguson2012predictive}. These instances, among others, stimulated the development of fairness-related research and policies, leading to the emergence of fairness-aware systems.

In recent years, fairness has evolved into a important part of responsible AI development. Researchers have proposed different metrics, methodologies, and regulatory frameworks to ensure that software systems adhere to ethical standards and do not perpetuate biases. Despite these efforts, ensuring fairness in software remains a complex challenge, especially as AI systems become more pervasive and intertwined with societal decision-making processes.


Although the community has made significant efforts to address fairness issues and propose solutions, many aspects remain unexplored \cite{barocas2023fairness}. This creates challenges for both researchers and practitioners in gaining a comprehensive understanding of the existing solutions for fairness in software, their key characteristics, and their potential for widespread industrial adoption.

Therefore, the primary objective of this study is to analyze the current state of the art in fairness solutions within software engineering. 
We aim to provide a comprehensive overview of existing research and practices related to fairness solutions. This analysis will not only highlight the advancements made in the field but also discuss gaps and challenges that remain unaddressed.

By having these studies thoroughly mapped and categorized, we can facilitate a deeper understanding among researchers and practitioners regarding the ongoing efforts in the realm of Software Fairness. This knowledge will help stakeholders to make informed decisions, foster collaboration, and inspire innovative solutions that promote equity and inclusivity in software systems. Furthermore, our findings will serve as a valuable resource for guiding future research directions and establishing best practices in the implementation of fairness solutions within software engineering.

To reach our objective, we employed a systematic mapping study, a research approach designed to objectively examining the full breadth of contributions in a particular field. In this study, we explored, categorized, and evaluated the current advancements in Software Fairness. We carefully selected 95 primary studies out of an initial pool of 1,722 relevant papers. Following this, we established a structured classification framework to systematically categorize findings in Software Fairness and applied it to these 1,722 studies. We then synthesized the data, offering a clear snapshot of the current landscape in Software Fairness. 

This study offers three primary contributions: (i) a guideline designed to classify, compare, and assess solutions, methods, and techniques specific to Software Fairness; (ii) a current overview of the state of the art in Software Fairness, providing insights and implications for future research; and (iii) a feasibility analysis in the industrial context.

Regarding the audience that this study aims to reach, we highlight two targets: (i) researchers looking to deepen their contributions in this field, and (ii) industry professionals aiming to understand current research on Software Fairness to selectively implement solutions that align with their organizational goals.

To the best of our knowledge, there is no systematic mapping study on Software Fairness that employs a framework-based methodology to systematically analyze studies and cluster them into specific aspects of research focus, research trends, and viability in the industrial context. This study is an extension of the paper titled ``Towards Fairness in AI: A Systematic Mapping Study on Software Engineering Solutions" \cite{nepomuceno2025fairness}. 

The paper is structured as follows: Section \ref{software_fairness} introduces foundational concepts in Fairness; Section \ref{study_design} outlines the study’s design, while Sections \ref{research_trends}, \ref{research_focus}, and \ref{viability_industrial_application} discuss the results in detail, offering a broader perspective and highlighting key implications for researchers and practitioners alike; Sections \ref{threats_validity} and \ref{related_works} cover potential limitations and related work, respectively; and, finally, Section \ref{conclusion} concludes the study with final thoughts.

\section{Background}
\label{software_fairness}

This section provides background information on the context of fairness. It is organized into two subsections: subsection 2.1 discusses definitions of fairness and the challenges associated with them, while subsection 2.2 introduces the concept of software fairness and related themes.

\subsection{Fairness}

Defining fairness in the context of software systems and responsible AI remains a challenging endeavor due to the absence of a universal definition fitting in diverses scenarios. One fundamental reason for this challenge is that fairness inherently depends on context, and what is considered fair in one scenario may be viewed as unfair in another. Thus, understanding the specific application context is essential before attempting to define fairness effectively.

The literature extensively illustrates the complexity and variety of fairness definitions by moving from one metric to another in search of a perfect fit for each unique situation \cite{caton}. Various studies highlight multiple quantitative definitions that attempt to capture fairness mathematically. Among these, the most widely recognized definitions include Equalized Odds, Equal Opportunity, and Demographic Parity \cite{mehrabi2021survey}.

Despite these comprehensive efforts, several researchers, \cite{corbett2023measure, green2018fair} argue that mathematical definitions often fail to encapsulate broader normative, social, economic, or legal understandings of fairness. Additionally, research indicates inherent limitations in simultaneously satisfying multiple fairness constraints, such as individual and group fairness, except under highly specific conditions \cite{kleinberg2016inherent}. Consequently, it becomes crucial to carefully choose fairness metrics based on the application context \cite{selbst2019fairness}. Moreover, current definitions may sometimes be counterproductive, potentially causing harm to sensitive groups over time, especially when measurement errors favor existing fairness criteria \cite{liu2018delayed}.

Given the many and often conflicting views on fairness, creating a single, universally applicable definition remains a major challenge. Overcoming this would greatly improve the consistency and comparability of fairness evaluations, helping advance the broader goal of building truly fair and responsible systems.

\subsection{Software Fairness}

The concept of Software Fairness offers a structured, objective approach to understanding fairness within the systems. By framing fairness as a core aspect of software behavior and as an essential property of software \cite{soremekun2022software}, we gain clearer insights into how biases may emerge at various stages of the software development life-cycle. Bias can be introduced during requirements gathering, system design, model development, data collection, or at any other point in the process. Therefore, a comprehensive approach to Software Fairness requires vigilant attention to each phase, ensuring that fairness is integrated from the ground up.

In software development, characteristics such as race or gender should not affect system outcomes such as eligibility for financial loans or access to job opportunities. However, real-world systems show that biases do occur: image search and translation engines have displayed gender stereotypes, while face detection and recognition tools have exhibited performance disparities across demographic groups \cite{barocas2023fairness}. These examples underscore the importance of prioritizing fairness in systems, not only for ethical and social reasons but to enhance the reliability and inclusivity of technology.

Failing to address this challenge early can create a growing backlog of fairness issues, a buildup of hidden costs and risks tied to unfair or biased outcomes. Addressing them later can become increasingly expensive, requiring redesigned user flows, model retraining, or even sweeping changes to data governance.

Achieving Software Fairness demands a multidisciplinary approach, incorporating technical solutions for bias detection and mitigation, policy-making for compliance, and ethical standards to guide responsible practices. By embracing fairness as a measurable and enforceable software property, we can create systems that are more transparent, accountable, and ultimately, equitable for all users.

This study takes a step toward addressing such biases by analyzing and categorizing scientific research on software fairness. By capturing and clustering studies in this field, we aim to deepen our understanding of the nature and challenges of fairness in systems. This structured view of current research can help identify both effective practices and gaps in addressing biases, guiding future efforts to develop more inclusive systems across various industries and applications.

\section{Study Design}
\label{study_design}

In this section, we describe the steps taken for the systematic mapping process. This approach allows for a clearer understanding of the motivations, criteria, and constraints that guided our methodology.  
We based our process on \cite{petersen2015guidelines}, which provided a structured foundation for systematically reviewing and categorizing relevant studies. Through this approach, we aim to ensure a comprehensive and transparent mapping of the field, highlighting key themes, gaps, and trends.

To guide our study, we address some research questions (RQs): \textit{ RQ1 -- What are the leading solutions proposed to achieve fairness in AI systems? RQ2 -- What are the primary topics within the field? RQ3 -- How are these proposed fairness solutions being integrated into software engineering? RQ4 -- What is the viability of the industrial context?}
To address these questions, we structure our research objectives into three main areas: identifying \textbf{research trends}, clarifying \textbf{research focus}, and assessing the \textbf{viability for industrial application}.

\subsection{Step 1: search and selection process}
\begin{figure*}
    \centering
    \includegraphics[width=\linewidth]{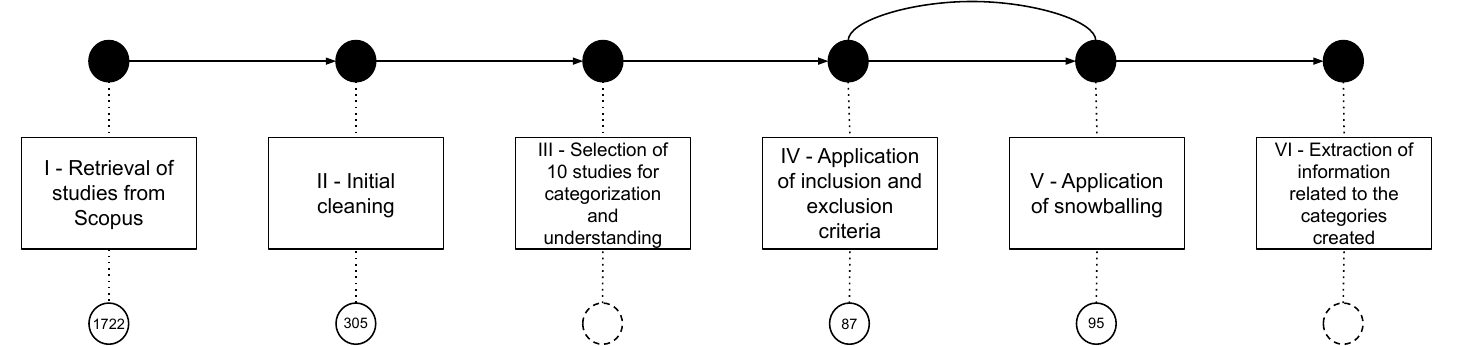}
    \caption{Flowchart of the entire process}
    \label{fig:flowchart}
\end{figure*}

To investigate the research trends and focus, and also the viability for industrial context, we conducted a systematic \textbf{search and selection} process, ensuring a comprehensive review of relevant literature. It corresponds to stage I in Figure~\ref{fig:flowchart}. We chose Scopus as our search database due to its extensive coverage across academic disciplines and its high-quality peer-reviewed content \cite{elsevier2016scopus}. Its robust data export capabilities and well-documented API made it an efficient choice for our analysis. Consequently, we relied exclusively on Scopus at all stages of our study.

We formulated the following search string to identify studies on fairness in the software domain, ensuring a broad and relevant collection:
\begin{center}
\textit{Software} AND \textit{Fairness}
\end{center}
We searched within the titles, abstracts, and keywords of the articles, limiting  the retrieval scope of our query to computer science to ensure domain relevance.

Database searches were run in August 2024 and updated in September 2025, yielding a total of 1,722 studies. For clarity, the counts presented at each phase in Figure 1 incorporate the September 2025 update. All studies and their associated metadata were exported in CSV format. This step produced our first artifact: a list of Scopus-indexed studies on software and fairness.

\subsection{Step 2: filtering}

Once we compiled the list of papers to be analysed, we conducted the first \textbf{filtering} step, screening titles and abstracts to exclude studies unrelated to software fairness, reducing the pool to 305 relevant studies (Stage II in Figure~\ref{fig:flowchart}).

\subsection{Step 3: keywording}
Then, using the \textbf{keywording} approach \cite{petersen2008systematic}, we identified the primary categories. First, we randomly selected ten papers to identify foundational keywords and concepts that helped us understand the thematic context of the studies (for example, keywords include: bias, bias mitigation, group fairness, individual fairness, processing stage, equal opportunity difference). Next, we grouped these keywords to create initial categories, forming the basis of our classification framework (e.g., categories such as “Fairness Metrics”, “Bias Type”, and “Dataset”). As we reviewed additional studies, we incorporated emerging concepts into the list of categories, refining and expanding it to ensure a comprehensive classification. These categories, which comprise our framework, are detailed in the "Results" section. We represented this process in stage III of Figure~\ref{fig:flowchart}.

\subsection{Step 4: selection criteria}
The next step involved applying our \textbf{selection criteria}, which were designed and discussed to align with our research objectives. The criteria, detailed in Table~\ref{tab:inclusion-exclusion-criteria}, allowed us to refine the studies according to the inclusion and exclusion rules, ensuring that the remaining studies provided meaningful insights aligned with our research objectives. This phase yielded 87 studies (stage~IV in Figure~\ref{fig:flowchart}).

\begin{table}
\setlength{\tabcolsep}{4.7pt}
\vspace{-3mm}
\caption{Inclusion and exclusion criteria}
\label{tab:inclusion-exclusion-criteria}
\begin{tabular}{+p{7cm}^p{7cm}}
\toprule
\rowstyle{\bfseries}
Inclusion Criteria & Exclusion Criteria \\
\midrule
I1: Studies presenting solutions specifically addressing Software Fairness. & 
E1: Studies that do not have fairness as a primary focus or that address it superficially. \\
\midrule
I2: Studies exploring challenges or problems in achieving Software Fairness. & 
E2: Secondary or tertiary studies, such as surveys or systematic reviews. \\
\midrule
I3: Empirical studies focused on analyzing fairness in software. & 
E3: Analyses or evaluations of fairness-related libraries or tools without original contributions. \\
\midrule
I4: Empirical studies focused on verifying and validating fairness in software. & 
E4: Case studies that fall outside the scope of our research objectives. \\
\midrule
I5: Peer-reviewed studies. & 
E5: Benchmarking studies that compare multiple models with datasets but lack substantive fairness-focused analysis. \\
\midrule
I6: Studies published in English.
& E6: Studies with fewer than 8 pages, to ensure sufficient depth and quality. \\
\bottomrule
\end{tabular}
\end{table}

\subsection{Step 5: snowballing}
We also applied the \textbf{snowballing} technique in an automated way to systematically broaden our range of relevant studies. Snowballing allowed us to identify additional sources by examining studies that cite or are cited by the initial set of studies. By the end of the process, we had 8 more scientific studies, totaling 95. To describe the process in more detail: starting with our 87 selected studies, we examined each one to check if the referenced or citing studies contained the terms software and fairness in their abstracts, title, and keywords. If a study included both terms and was not a duplicate, it was added to our dataset (our set of selected studies) and subjected to the previously mentioned screening steps. 

The snowballing technique was conducted both forward and backward to a depth of three levels. For instance, our initial list (List A) comprised 87 studies. After the first round of snowballing, if 3 new relevant studies were identified, they were added to List A. We then applied the snowballing technique to these 3 new studies, initiating the second round. If this round yielded 2 additional studies, they were incorporated into List A, and we proceeded to the third round of snowballing. Finally, if the third round identified 1 more relevant study, we completed the snowballing process at this third and final level.
This snowballing process was automated using the Scopus API and the Pybliometrics library. We can see the snowballing represented by stage V in Figure~\ref{fig:flowchart}.

\subsection{Step 6: information extraction}
\begin{figure*}
    \centering
    \includegraphics[width=1\linewidth]{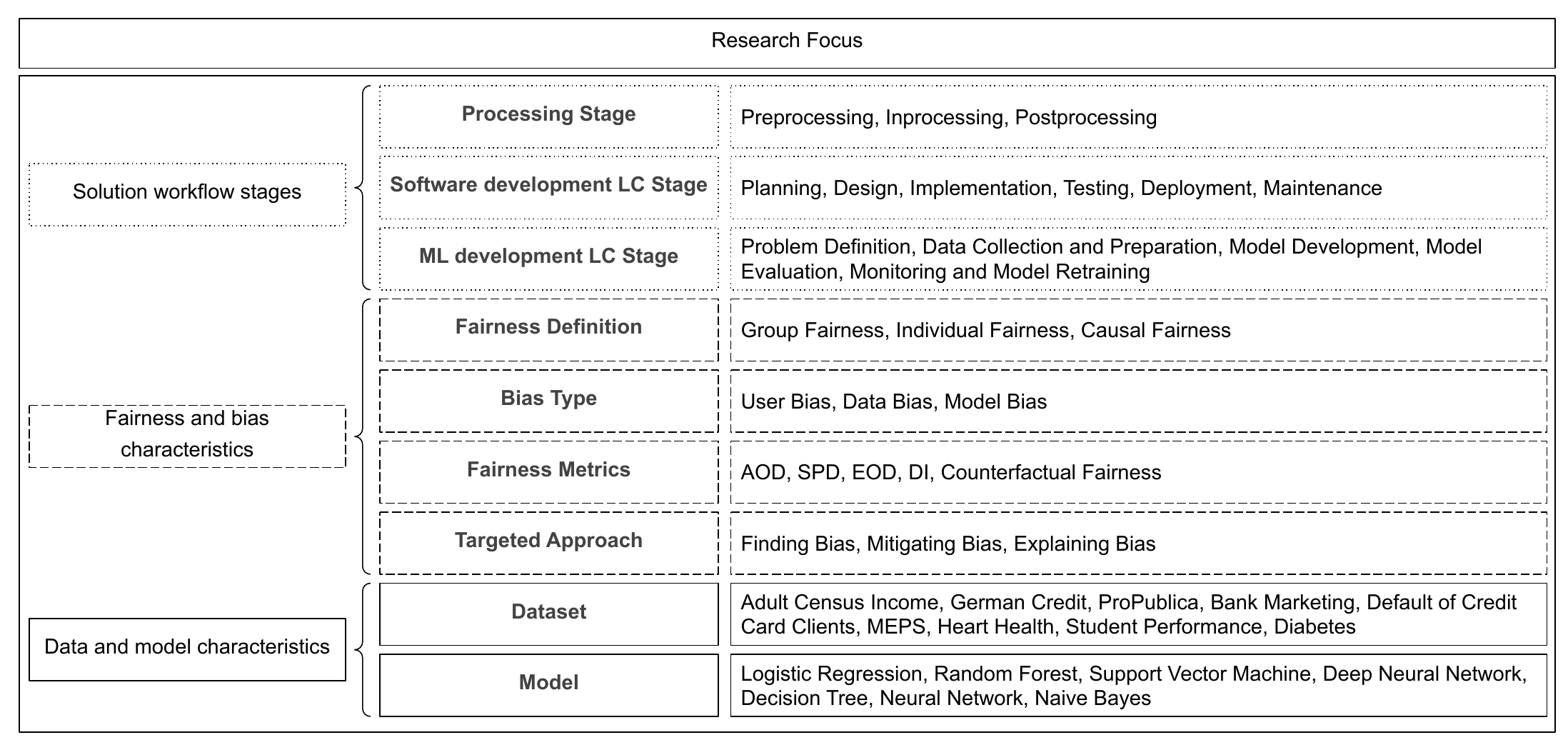}
    \caption{Research focus detailed}
    \label{fig:researchfocus}
\end{figure*}

After identifying all relevant studies, we completed the process of finding and selecting the studies. In total, we found 95 studies. With this set of studies, we proceeded to the next phase, which involved extracting the necessary information. Thus, we began the \textbf{information extraction} process (stage VI in Figure~\ref{fig:flowchart}). During this phase, we read each selected study to extract the necessary information. 
The extracted information can be summarized into the following three groups: 

\begin{itemize}
    \item Research Trends: we analyzed the publication year, venue type, trends in problems and solutions, geographic distribution (publication country), and the most commonly mentioned approaches;
    \item Research Focus: we examined the solution workflow stages, fairness and bias characteristics, as well as dataset and model characteristics. It is better represented in Figure~\ref{fig:researchfocus};
    \item Viability for Industrial Application: this aspect comprises four factors: 1) technology readiness level (TRL); 2) industrial and academic engagement; 3) tool support; and 4) open availability to the community.
\end{itemize}

After accomplishing the categorization of all studies, we synthesized the categories into comprehensive tables, allowing us to organize information systematically for deeper analysis and discussion (we can see the tables in the Sections \ref{research_trends}, \ref{research_focus}, and \ref{viability_industrial_application} -- Results). The tables consist of three columns: the subcategory name, the amount of studies, and the code number corresponding to each study. Exception are Tables~\ref{tab:solutiontrend} and \ref{tab:dataset}, where the third column indicates the source of the aproach and respective dataset. The mapping between code numbers and paper titles is available in the document on Zenodo \cite{nepomuceno_2025_17435899}.

To conclude this section, it is important to highlight that the methodology presented facilitated a thorough and systematic review ensuring that our research includes all pertinent studies on Software Fairness and provides a solid foundation for further insights and discussions.

\section{Results -- Research Trends}
\label{research_trends}

The motivation behind analyzing research trends is to understand what topics researchers are discussing and what solutions they find most interesting in the field of fairness. This helps to gain a deeper understanding of the current state of the area. To achieve this, we examined various aspects, including (1) publication year and venue type; (2, 3) key trends in problems and solutions; (4) geographic distribution (publication country); and (5) the most commonly mentioned approaches. We will examine each of these aspects in detail.

\subsection{Publication Year and venue type}

\begin{figure}
    \centering
    \includegraphics[width=0.7\linewidth]{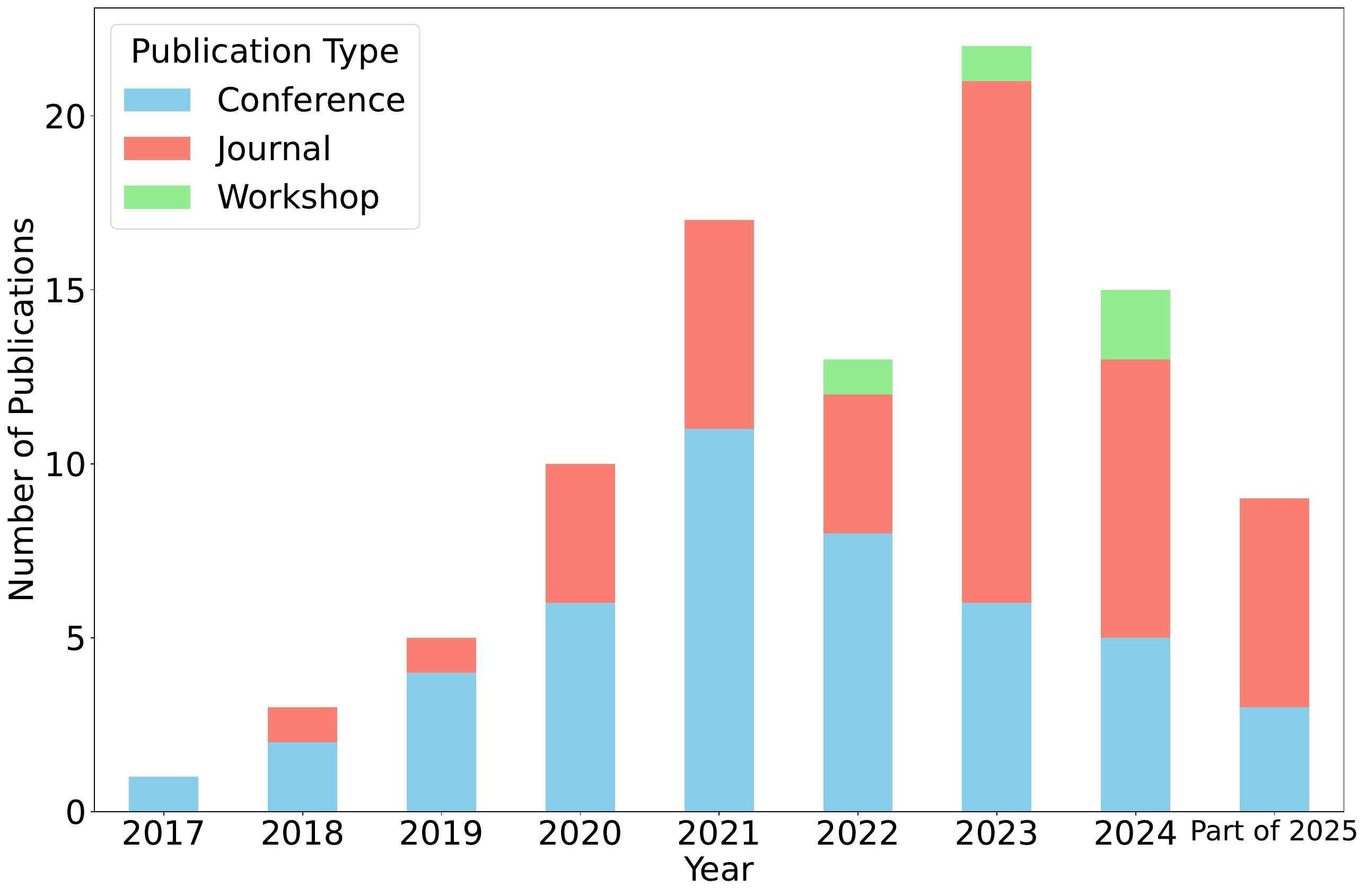}
    \caption{Distribution of primary studies by type of publication over the years}
    \label{fig:publicationsl}
\end{figure}

Figure~\ref{fig:publicationsl} illustrates the growth trajectory of scientific publications related to Software Fairness over the past years. Initially, there was a modest increase in publications, indicating early interest in the topic. However, recent years have shown a significant rise in research output, reflecting the increased societal usage of systems and the growing awareness of fairness in technologies. This trend likely stems from the widespread integration of software and AI software in everyday life, making Software Fairness an essential area of study. Notably, the first studies explicitly using terms like “Software Fairness” or directly addressing fairness in software emerged around 2017, marking a foundational period in this field \cite{galhotra2017fairness}.

We also observe the distribution of publications per year and by venue type, providing further insights into the maturity of the discussion surrounding software fairness. Specifically, the set of studies analyzed includes 4 workshop papers, 45 journal articles, and 46 conference papers, which suggests that while the field is relatively young, it has attracted substantial academic attention across a range of venues. This variety in publication types demonstrates a solid and evolving foundation for the topic, with discussions taking place at multiple levels of academic discourse.

Therefore, given the growth we have already achieved and the solid foundation demonstrated by the variety of venue types in recent years, we can say that the area will continue having attention in the coming years.

\subsection{Trends in Problems (Problems Categorized)}

We present the categorization of the issues identified in the studies. Our goal is to establish a clearer mapping between problems and their corresponding solutions, thereby enhancing the understanding of the overall context within the field. To achieve this, we have organized the problems into the following categories:

\begin{itemize}

\item \textbf{Verification and validation issues}: these issues arise due to the unreliable behavior of systems under varying conditions. Ensuring that systems behave as intended in all scenarios is challenging, which can lead to potential failures or unreliable outcomes.

\item \textbf{Poor data quality, bias, or insufficient data (data issues)}: data related issues encompass poor data quality, insufficient data, or biased datasets leading to flawed or unreliable software systems.

\item \textbf{Models failing to generalize or overfitting (model issues)}: model related issues often are models that fail to generalize, overfit, or underperform due to inherent limitations in their design or training.

\item \textbf{Misalignment with real-world requirements and design}: design and development challenges lead to complexity in designing systems that align with real-world requirements and constraints. This complexity can result in mismatched solutions.


\item \textbf{Tradeoff issues}: optimization and trade-offs issues is the inability to simultaneously optimize multiple competing objectives (e.g., performance vs. fairness), leading to suboptimal outcomes.

\item \textbf{Algorithmic issues}: algorithmic fairness is a critical concern, as biases in systems can lead to unfair or discriminatory outcomes for certain groups.


 \item \textbf{Lack of explainability in AI decisions}: lack of explainability in models, making it difficult to understand or trust their decisions.
 
\end{itemize}

According to Table~\ref{tab:problems}, data and model issues, along with verification and validation issues, are the most frequently addressed types of problems. In contrast, the explainability of systems remains the least prioritized area. This suggests that researchers and practitioners in the field tend to prioritize data quality and model performance, as well as to ensure the correctness of systems rather than explaining them.

\begin{table}[t]
\setlength{\tabcolsep}{4.7pt}
\vspace{-3mm}
\caption{Problems categorized}
\label{tab:problems}
\begin{tabular}{+c^c^c}
\toprule
\rowstyle{\bfseries}
Problem Categories & Amount of Studies & Studies \\
\midrule
\makecell{Data and model issues} & 36 & \makecell{S1, S3, S5, S6, S8, S14, S19, S21, S25, S26, S27, S28, S29, S34, S38, S40, S42, S43, S51, \\ S53, S55, S56, S57, S62, S72, S75, S76, S77, S82, S85, S86, S89, S90, S91, S92, S93} \\
\midrule
Verification and validation issues & 34 & \makecell{S2, S7, S9, S10, S11, S12, S13, S15, S17, S18, S20, S22, S23, S30, S31, S32, S33, S35, \\ S45, S46, S48, S51,  S58, S63, S64, S73, S77, S78, S83, S84, S87, S88, S94, S95} \\
\midrule

\makecell{Misalignment with real-world \\ requirements and design} & 15 & \makecell{S4, S19, S28, S37, S39, S41, S47, S55, S60, S66, S67, S68, S70, S71, S79} \\
\midrule
Algorithmic issues & 15 & \makecell{S24, S29, S34, S49, S50, S52, S59, S60, S65, S69, S72, S74, S76, S80, S81 } \\
\midrule
Tradeoff issues & 10 & \makecell{S6, S16, S24, S36, S37, S38, S44, S54, S59, S75} \\
\midrule
\makecell{Lack of explainability} & 4 & \makecell{S52, S61, S74, S77} \\
\bottomrule
\end{tabular}
\end{table}

\subsection{Trends in Solutions (Solution Approach)}

We categorized the different solutions into six categories:

\begin{itemize}
\item \textbf{Empirical Analyses}: studies that evaluate the effectiveness or biases of existing systems through experiments, observational data, or real-world testing. These analyses provide insights into how current systems perform in practice and identify areas for improvement.

\item \textbf{Metric Proposals}: the development of new metrics designed to measure specific aspects of systems, such as fairness, bias, transparency, and accountability. These proposals offer fresh perspectives and more precise tools for evaluating models.

\item \textbf{Tool Developments}: the creation of software tools that assist in tasks such as fairness auditing, model interpretation, and bias detection. These tools aim to make AI systems more understandable and accessible to non-expert users, enabling more effective oversight and compliance.

\item \textbf{Frameworks}: structured approaches, guidelines, or methodologies that provide adaptable, step-by-step processes for addressing complex challenges. These frameworks help ensure that systems are developed and evaluated consistently and systematically, promoting best practices in the field.

\item \textbf{Methods and Algorithms}: novel techniques or algorithms designed to solve specific technical problems, such as improving fairness, optimizing performance, or enhancing transparency. These methods address the core computational and theoretical challenges in software fairness.

\item \textbf{Testing and Verification Methods}: techniques and processes used to assess the reliability, fairness, and correctness of models. This includes generating tests to validate the performance of systems or conducting verification checks to ensure they meet predefined standards for safety, fairness, and accuracy.

\end{itemize}

As shown in Table~\ref{tab:test}, the most frequent solutions are Methods \& Algorithms, and Frameworks, highlighting a focus on practical and implementable solutions in Software Fairness research. 
In contrast, Metric Proposals is the least explored area in the identified studies. This raises two potential interpretations regarding metrics:
\begin{enumerate}
\item A gap in the area regarding the development of metrics specifically designed to evaluate fairness in systems;

\item Alternatively, it could indicate that existing metrics are already sufficiently comprehensive in capturing the concept of fairness.
\end{enumerate}

\begin{table}
\setlength{\tabcolsep}{4.7pt}
\vspace{-3mm}
\caption{Solutions categorized}
\label{tab:test}
\begin{tabular}{+c^c^c}
\toprule
\rowstyle{\bfseries}
Solution Approach Categories & Amount of Studies & Studies \\
\midrule
Methods \& Algorithms & 29 & \makecell{S1, S3, S5, S6, S12, S14, S16, S21, S24, S26, S31, S38, S40, S45, S49, S50, S54, \\ S60, S62, S76, S84, S87, S88, S89, S90, S91, S92, S93, S94} \\
\midrule
Frameworks (Approach) & 23 & \makecell{S4, S7, S19, S28, S34, S39, S43, S47, S52, S53, S56, S59, S66, S69, S70, S71, S72, \\ S73, S75, S77, S78, S80, S85} \\
\midrule
\makecell{Testing \& Verification \ Methods} & 13 & \makecell{S11, S13, S17, S18, S20, S23, S30, S33, S35, S51, S64, S83, S95} \\
\midrule
Empirical Analyses & 12 & \makecell{S8, S9, S25, S36, S37, S41, S44, S46, S65, S67, S68, S79} \\
\midrule

Tool Developments & 10 & \makecell{S2, S15, S22, S29, S32, S48, S58, S61, S81, S86} \\
\midrule
Metric Proposals & 8 & \makecell{S10, S27, S42, S55, S57, S63, S74, S82} \\
\bottomrule 
\end{tabular}
\end{table}

This discrepancy invites further exploration into whether existing fairness metrics are adequate or if additional work is required to create more nuanced or specialized measures for evaluating AI systems.

In Table~\ref{tab:problemxsolution}, we can observe the relationship between the problems and the solutions identified in the analyzed studies. It allows us to map the types of solutions applied to specific problems and gain insight into the prevailing trends in research approaches.
According to the table, ``data and model issues'' and ``verification and validation issues'' are the only types of problems that have all kinds of solutions. ``Frameworks'' appear as a solution for all problem categories, suggesting that researchers primarily address fairness challenges through frameworks. This indicates that frameworks are seen as a key approach to tackling fairness-related issues.
Additionally, we observe that issues related to validation, data, and models are receiving attention across all solution categories, highlighting their significance in fairness research.

\begin{table}[]
\setlength{\tabcolsep}{4.7pt}
\vspace{-3mm}
\caption{Problems vs solutions}
\label{tab:problemxsolution}
\begin{tabular}{+l^c^c^c^c^c^c}
\toprule
\rowstyle{\bfseries}
 & \makecell{Frameworks \\ (Approaches)} & \makecell{Methods \\ \& Algorithms} & \makecell{Empirical \\Analyses} & \makecell{Testing \& \\ Verification Methods} & \makecell{Tool \\Developments} & \makecell{Metric \\Proposals} \\
\midrule
Issue 1 & \makecell{S19, S28, S43, S53, \\ S56, S75, S34, \\ S72, S77, S85} & \makecell{S1, S3, S5, S14, \\ S21, S40, S6, S38, \\ S62, S76, S89, S90, \\ S91, S92, S93} & \makecell{S8, S25, S26} & \makecell{S51} & \makecell{S29, S86} & \makecell{S27, S42, S55, S57, \\ S82} \\
\midrule
Issue 2 & \makecell{S7, S73, S77, S78} & \makecell{S12, S31, S45, S84, \\ S87, S88, S94} & \makecell{S9, S46} & \makecell{S11, S13, S17, S18, \\ S20, S23, S30, S33, \\ S35, S51, S64, S83, \\ S95} & \makecell{S2, S15, S22, S32, \\ S48, S58} & \makecell{S10, S63} \\
\midrule
Issue 3 & \makecell{S4, S19, S28, S39, \\ S47, S66, S70, S71} & \makecell{S60} & \makecell{S37, S41, S67, S68, \\ S79} & \makecell{} & \makecell{} & \makecell{S55} \\
\midrule
Issue 4 & \makecell{S34, S52, S59, S69, \\ S72, S80} & \makecell{S24, S49, S50, S60, \\ S76} & \makecell{S65} & \makecell{} & \makecell{S29, S81} & \makecell{S74} \\
\midrule
Issue 5 & \makecell{S59, S75} & \makecell{S6, S16, S24, S38, \\ S54} & \makecell{S36, S37, S44} & \makecell{} & \makecell{} & \makecell{} \\
\midrule
Issue 6 & \makecell{S52, S77} & \makecell{} & \makecell{} & \makecell{} & \makecell{S61} & \makecell{S74} \\
\bottomrule
\multicolumn{7}{l}{\textbf{Issue Descriptions:}} \\
\multicolumn{7}{l}{Issue 1: Data and model issues} \\
\multicolumn{7}{l}{Issue 2: Verification and validation issue} \\
\multicolumn{7}{l}{Issue 3: Misalignment with real-world requirements and design} \\
\multicolumn{7}{l}{Issue 4: Algorithmic issues} \\
\multicolumn{7}{l}{Issue 5: Tradeoff issues} \\
\multicolumn{7}{l}{Issue 6: Lack of explainability} \\
\bottomrule
\end{tabular}
\end{table}

\subsection{Geographic Distribution}

\begin{figure}
    \centering
    \includegraphics[width=1\linewidth]{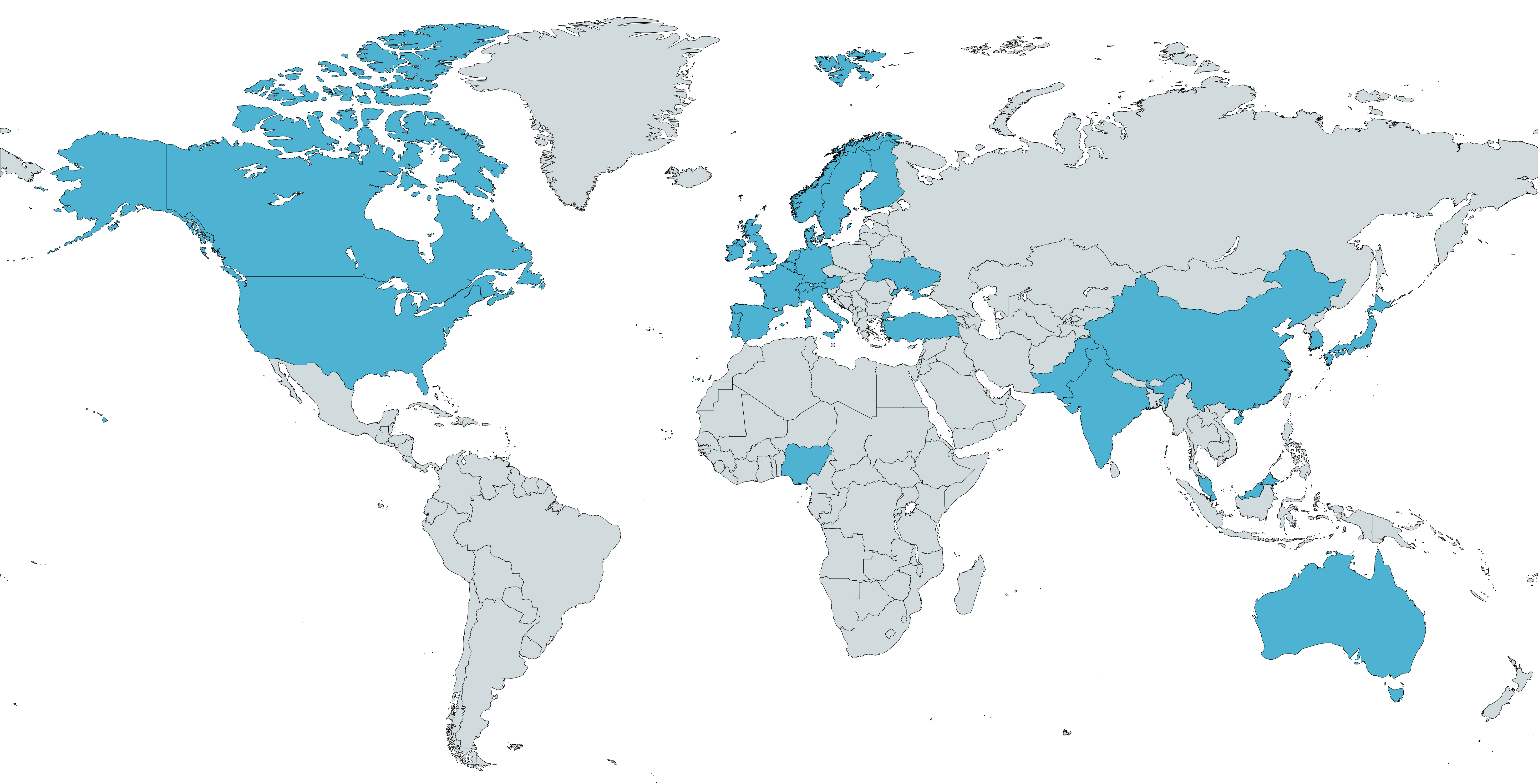}
    \caption{Distribution of countries in AI fairness research}
    \label{fig:countries}
\end{figure}

By examining the global distribution of research on software fairness (Figure~\ref{fig:countries}), it reveals that researches are largely concentrated in Global North \cite{UNCTAD2025}, in socioeconomics terms, with the exceptions of some countries. While it is difficult to pinpoint the exact reasons for this disparity, it may be linked to the concentration of resources, funding, and institutional focus on technologies in these regions. However, fairness is a universal concept that transcends cultural and geographical boundaries, and its interpretation can vary significantly across different societies.

To truly address fairness in a global context, it is essential to incorporate diverse cultural perspectives. A more inclusive approach would not only enrich the discourse but also ensure that the development of AI and related technologies is equitable and representative of the world’s varied values and norms. Unfortunately, we are still far from achieving this level of diversity in research and practice. Moving forward, fostering collaboration and amplifying voices from underrepresented regions will be crucial in creating a more balanced and holistic understanding of fairness.


\subsection{Most Commonly Mentioned Approaches}

In this analysis, we examined the fairness solutions that are frequently mentioned or used for comparison. It is interesting to observe the trending approaches that researchers and practitioners discuss the most and incorporate as core methods. Table~\ref{tab:solutiontrend} presents the ten most commonly mentioned fairness solutions, with ``Reweighing" being the most frequently cited. Understanding the most referenced approaches not only highlights where the community is focusing its attention and drawing inspiration for new developments but also provides valuable insights into which directions we should explore, or reconsider, when designing new solutions.

\begin{table}
\setlength{\tabcolsep}{4.7pt}
\vspace{-3mm}
\caption{Most commonly mentioned approaches}
\label{tab:solutiontrend}
\begin{tabular}{+c^c^c}
\toprule
\rowstyle{\bfseries}
Most commonly mentioned approach & Amount of Studies & Source \\
\midrule
Reweighing & 13 & Ref: \cite{li2022achieving} \\
\midrule
AEQUITAS & 11 & Ref: \cite{saleiro2018aequitas} \\
\midrule
Fair-SMOTE & 10 & Ref: \cite{chakraborty2021bias} \\
\midrule
Symbolic Generation (SG) & 7 & Ref: \cite{aggarwal2019black} \\
\midrule
\makecell{Adversarial Debiasing} & 7 & Ref: \cite{zhang2018mitigating} \\
\midrule
Reject Option Classification & 6 & Ref: \cite{herbei2006classification} \\
\midrule
Adversarial Discrimination Finder & 5 & Ref: \cite{zhang2020white} \\
\midrule
\makecell{Prejudice Remover \ Regularizer} & 4 & Ref: \cite{kamishima2012fairness} \\
\midrule
Fairway & 4 & Ref: \cite{chakraborty2020fairway} \\
\midrule
Themis & 3 & Ref: \cite{angelopoulos2009themis} \\
\bottomrule 
\end{tabular}
\end{table}

\begin{tcolorbox}[colframe=black!50, colback=white, sharp corners, boxrule=0.5pt, width=\linewidth]
\textit{Main findings:}
\begin{itemize}
    \item There is a growing tendency for the Software Fairness field to expand further in the coming years.
    \item When discussing fairness, people often perceive it as an issue of unreliable systems that do not perform as intended, requiring validation in terms of fairness. This is also seen as a data and model problem, which is primarily addressed through procedural frameworks.
    \item The fact that metrics have the fewest studies associated with them suggests a gap in the field.
    \item There is a low representation of fairness research from the global south. (The terms global south and global north represent socioeconomic and political aspects rather than strictly geographic divisions. The global south remains underrepresented compared to the global north \cite{unctadstat_classifications}).
    \item Research on software fairness emphasizes practical and implementable solutions.
    \item Data and model issues, along with Verification and validation challenges, are central concerns in fairness research.
\end{itemize}
\end{tcolorbox}

\section{Results -- Research Focus}
\label{research_focus}

This section presents the findings related to one of the central objectives of this study: identifying the primary areas of focus in current research on Software Fairness and examining how these areas have been incorporated into Software Engineering (SWE) principles and practices. To systematically address these aspects, we organized our analysis into three key dimensions: (1) Solution Workflow Stage, (2) Fairness and Bias Characteristics, and (3) Dataset and Model Characteristics. These subgroups provide a structure for exploring the evolving landscape of Software Fairness research and its integration into SWE.

\subsection{Solution Workflow Stage}

This subgroup examines the different stages to build an AI or machine learning solution. These stages are critical to understanding how AI/ML models are developed, tested, deployed, and maintained. By breaking down the life-cycle into distinct steps, we can better assess how systems and fairness are structured and how they interact with software engineering practices. The workflow analyzed here will be (i) processing stages, (ii) software development life-cycle, and (iii) ML development life-cycle.

\subsubsection{Processing Stage}

Processing stage refers to the three phases in the data processing pipeline:

\begin{itemize}
  
\item Pre-processing: this is the phase where data is prepared before feeding it into a model \cite{mehrabi:2021}.  
It includes data preparation tasks such as retrieval, cleaning, formatting, and elaboration. In this work, we consider any step influencing model fairness before training as part of the pre-processing stage.

\item In-processing: this is the stage where the model is trained or the actual prediction/learning happens. It includes the application of algorithms, training of models, hyperparameter tuning, etc \cite{mehrabi:2021}.

\item Post-processing: this phase includes tasks after the model has been deployed or predictions are made. It might involve steps like evaluation, interpretation, and model refinement or explaining the model's behavior to ensure it is functioning as expected \cite{mehrabi:2021}. In this mapping, we include any study addressing the solution after the training phase.

\end{itemize}

 Through the results, we can see a significant proportion of the studies (45 out of 95) focuses on post-processing techniques, suggesting that many fairness interventions are applied after the model is trained. This might indicate that post-processing methods are often favored for their flexibility in adjusting model outputs without changing the training process or model architecture.
Although post-processing has the highest count, the other stages, pre-processing (33/95) and in-processing (30/95), are also well-represented, showing a relatively balanced approach to addressing fairness across the data processing pipeline. This balance may imply a comprehensive effort in the community to address fairness throughout the pre-processing and in-processing. Emphasizing fairness from the outset ensures that issues are proactively managed, rather than relying solely on corrections after the training phase.

\begin{table}
\setlength{\tabcolsep}{4.7pt}
\vspace{-3mm}
\caption{Processing stage}
\label{table:processing}
\begin{tabular}{+c^c^c}
\toprule
\rowstyle{\bfseries}
Processing Stage & Amount of Studies & Studies \\
\midrule
Pre-processing & 33 & \makecell{S1, S2, S3, S4, S5, S8, S14, S17, S19, S21, S25, S26, S27, S28, S36, S40, S41, S42, S43, S44, \\ S47, S55, S56, S60, S62, S66, S75, S82, S85, S86, S87, S88, S93} \\
\midrule
In-processing & 30 & \makecell{S1, S6, S8, S14, S15, S16, S28, S29, S31, S34, S35, S36, S37, S41, S44, S45, S46, S49, S50, \\ S51, S52, S54, S60, S72, S75, S76, S77, S81, S90, S91, S92} \\
\midrule
Post-processing & 45 & \makecell{S5, S7, S8, S9, S10, S11, S12, S13, S15, S18, S20, S22, S23, S24, S28, S30, S32, S33, S36, S38, \\ S39, S41, S44, S46, S48, S52, S58, S59, S60, S61, S64, S65, S69, S73, S74, S77, S78, S80, S83, \\ S84, S87, S89, S94, S95} \\
\bottomrule 
\end{tabular}
\end{table}

\subsubsection{Software development LC Stage}

This analysis refers to the software development life-cycle steps. It is a general reference to the stages a system or AI software project goes through, from the initial requirement gathering, design, development, testing, deployment, and maintenance \cite{standard2008systems}.

The majority of studies are concentrated in the Implementation (46 out of 95) and Testing (38 out of 95) stages. This emphasis suggests that a significant portion of fairness efforts are centered around building and verifying systems. Researchers and practitioners may focus on these stages because they offer opportunities to directly apply fairness metrics, tune model parameters, and observe fairness outcomes, which are critical for ensuring a fair system before it is deployed.
Only 7 studies address Planning, indicating a relatively lower focus on fairness in early phases. This could imply that fairness considerations are often delayed until later in the development process. Integrating fairness considerations into the Planning stage, however, could proactively shape fairer approaches before development begins, potentially reducing fairness issues down the line. This underrepresentation could be an area for future focus, as early-stage interventions might lead to more general fairness outcomes.

The Deployment and Maintenance stages have 7 and 9 studies, respectively, which is comparatively low. Low attention in these stages may highlight an area for improvement, as fairness monitoring and adjustments in real-time environments are crucial for sustained system fairness over time.
The presence of studies across all stages suggests that fairness is being addressed in a life-cycle-oriented manner, yet the uneven distribution indicates that some stages might benefit from more robust methodologies or frameworks for fairness. Integrating fairness practices across stages, from planning to maintenance, could create a more cohesive approach to fairness, potentially reducing bias introduced or amplified at each stage.
Given the varying focus on each stage, there may be an opportunity to develop standardized practices for fairness at each life-cycle phase, to create a more consistent approach to fairness.



\begin{table}[ht]
\setlength{\tabcolsep}{4.7pt}
\vspace{-3mm}
\centering
\caption{Software development LC stage}
\label{tab:swe_lc_stage}
\begin{tabular}{+c^c^c}
\toprule
\rowstyle{\bfseries}
Software Development LC Stage & Amount of Studies & Studies \\
\midrule
Planning & 7 & \makecell{S28, S36, S41, S56, S67, S68, S87} \\
\midrule
Design & 24 & \makecell{S1, S4, S17, S19, S27, S28, S36, S41, S47, S53, S55, S56, S60, S62, S66, \\ S67, S68, S70, S71, S79, S82, S85, S86, S87} \\
\midrule
Implementation & 46 & \makecell{S1, S3, S5, S6, S8, S9, S16, S21, S25, S26, S27, S28, S29, S31, S34, S36, S37, \\ S38, S40, S41, S42, S43, S44, S45, S46, S49, S50, S51, S52, S54,  S60, S61, \\ S62, S72, S75, S76, S77, S80, S81, S82, S84, S88, S90, S91, S92, S93} \\
\midrule
Testing & 38 & \makecell{S2, S7, S11, S13, S14, S18, S20, S22, S23, S24, S26, S28, S30, S31, S32, S33, \\ S35, S36, S39, S41, S46, S48, S52, S58, S59, S60, S64, S65, S73, S74, \\ S77, S78, S80, S83, S87, S89, S94, S95} \\
\midrule
Deployment & 7 & \makecell{S12, S28, S36, S41, S65, S69, S77} \\
\midrule
Maintenance & 9 & \makecell{S12, S15, S23, S28, S36, S41, S58, S59, S77} \\
\bottomrule
\end{tabular}
\end{table}

\subsubsection{ML development LC Stage}

The ML development life-cycle encompasses the complete process of building an ML system. The results follow the behavior of the previous analysis, highlighting that fairness efforts are well represented in the technical stages, particularly in Monitoring and Model Development, with 34 out of 95 and 31 out of 95, respectively. However, there is a critical opportunity to strengthen fairness considerations in the earlier stage of Problem Definition (6 out of 95). Setting clear fairness goals and ensuring unbiased data inputs from the outset can establish a stronger, fairer foundation, potentially reducing the need for extensive AI fairness interventions later in the development life-cycle.

\begin{table}
\setlength{\tabcolsep}{4.7pt}
\vspace{-3mm}
\caption{ML development LC stage}
\label{tab:ml_lc_stage}
\begin{tabular}{+c^c^c}
\toprule
\rowstyle{\bfseries}
ML LC Stage & Amount of Studies & Studies \\
\midrule
Problem Definition & 6 & \makecell{S17, S36, S41, S47, S53, S56} \\
\midrule
\makecell{Data Collection and Preparation} & 25 & \makecell{S1, S3, S5, S8, S9, S19, S21, S25, S26, S27, S28, S36, S40, S41, S42, S43, \\ S55, S56, S62, S66, S75, S82, S85, S86, S88} \\
\midrule
Model Development & 31 & \makecell{S6, S8, S9, S16, S28, S29, S31, S34, S36, S37, S38, S41, S44, S45, S46, S49, S50, \\ S51, S52, S54, S60, S72, S75, S76, S77, S81, S84, S90, S91, S92, S93} \\
\midrule
Model Evaluation & 28 & \makecell{S5, S8, S11, S28, S29, S31, S34, S36, S37, S38, S41, S44, S45, S46, S49, S50, \\ S51, S52, S54, S60, S65, S72, S75, S77, S81, S87, S94, S95} \\
\midrule
\makecell{Monitoring and Model Retraining} & 34 & \makecell{S2, S7, S12, S13, S14, S15, S18, S20, S22, S23, S24, S28, S30, S31, S32, S33, \\ S35, S36, S39, S41, S48, S52, S58, S59, S60, S61, S64, S65, S69, S73, S77, \\ S78, S80, S83} \\
\bottomrule
\end{tabular}
\end{table}

\subsection{Fairness and Bias Characteristics}

In this subsection, we explore (i) key fairness concepts, (ii) the types of bias that can impact the systems outcomes, (iii) commonly used fairness metrics, and (iv) we examine the efforts of researchers to understand and address fairness issues, highlighting strategies for dealing with bias, that is, the targeted approach. This exploration aims to provide a comprehensive view of how fairness considerations are used in developing solutions to ensure ethically sound and socially responsible systems.

\subsubsection{Fairness Definition}

We observed how the literature on Software Fairness addresses fairness definitions, just few studies bring a clear statement about a definition, and most sources classify fairness into two main types: group fairness, and individual fairness, either explicitly or using representative metrics for this.

As we can see in Table~\ref{tab:fairness_definition}, with 56 out of 95 papers focusing on group fairness, this concept dominates fairness discussions. Group fairness often involves ensuring that groups (like demographics or protected classes) are treated equitably, highlighting the field's concern for addressing biases across different groups rather than on an individual level. 36 out of 95 explore individual fairness. This suggests a significant but less prominent interest in ensuring fairness at the individual level, where each person is treated similarly to others with comparable attributes. Lastly, only two papers address dependency or causal fairness, a concept that aims to account for the causal relationships and dependencies that might lead to biased outcomes. This scarcity indicates that the causal fairness approach is still emerging or underexplored in fairness research.

We can speculate that group fairness gets more attention because it is often tied to systemic discrimination and social justice concerns. When people discuss fairness, they usually think about large-scale disparities, like race, gender, or socioeconomic status, because these issues affect entire communities, not just individuals. This makes group fairness easier to study, measure, and act upon at a policy level. But, one downside is that group fairness can sometimes overlook individual fairness. Two individuals from different groups might have very different needs or experiences, but if we only focus on ensuring group-level parity, we might fail to address specific cases of unfairness at the individual level.

\begin{table}
\setlength{\tabcolsep}{4.7pt}
\vspace{-3mm}
\caption{Fairness definition}
\label{tab:fairness_definition}
\begin{tabular}{+c^c^c}
\toprule
\rowstyle{\bfseries}
Fairness Definition & Amount of Studies & Studies \\
\midrule
Group Fairness & 56 & \makecell{S1, S2, S3, S4, S5, S6, S8, S9, S10, S12, S14, S16, S19, S21,  S24, S25, S26, \\ S27, S29, S31, S36, S38, S40, S41, S42, S43, S44, S46, S48, S49, S52, S53, \\ S54, S55, S56, S57, S59, S60, S63, S64, S65, S67, S68, S72, S73, \\ S76, S77, S81, S84, S85, S87, S88, S89, S91, S92, S93} \\
\midrule
Individual Fairness & 36 & \makecell{S2, S3, S4, S5, S7, S9, S10, S13, S15, S17, S18, S19, S22, S23, S30, \\ S32, S33, S34, S35, S36, S41, S45, S50, S58, S74, S75, S77, S80, \\ S81, S83, S86, S87, S90, S93, S94, S95} \\
\midrule
Dependency Fairness (Causal Fairness) & 2 & \makecell{S11, S20} \\
\bottomrule 
\end{tabular}
\end{table}

\subsubsection{Bias Type}

Considering the definitions provided in \cite{mehrabi2021survey} and our observations during the mapping, we have categorized the different types of bias into three distinct groups:

\begin{itemize}

\item \textbf{User Bias}: This type of bias arises from human influence during the software engineering process, often introduced early in stages such as requirements gathering, design, or user interaction design. User bias can stem from the decisions, assumptions, or perspectives of those involved in the software's creation, potentially embedding subjective or unintended biases into the system.

\item \textbf{Data Bias}: Data bias occurs when the dataset used to train or evaluate a model is itself biased. This can happen if the dataset is unrepresentative, incomplete, or reflects historical prejudices. As a result, biased data can propagate discriminatory outcomes in the model, leading to unfair or unequal treatment.

\item \textbf{Model Bias}: Model bias is introduced when the model’s structure, algorithms, or learning processes inherently produce biased outcomes. This type of bias may arise from the model's design choices, optimization methods, or regularization techniques that amplify certain patterns or ignore others. Model bias directly affects output, creating the potential for unfair treatment or misclassification.
\end{itemize}
These categories highlight how bias can be introduced at multiple stages of software development, from human decisions in design (user bias) to data issues (data bias) and to the model's learning process (model bias).

Of the 95 papers analyzed, 64 focus on model bias, making it the dominant area of interest in fairness research. Model bias arises from algorithms, architectures, or learning processes within the model and is considered a critical source directly influencing outcomes. Addressing model bias is essential because it can have a significant impact through direct interventions.

Data bias is discussed in 34 papers, indicating substantial concern over fairness issues stemming from training datasets that may be incomplete, unrepresentative, or reflective of historical inequities. Biased data can propagate unfairness into models, highlighting the need for better data collection, preprocessing, and balancing techniques.

Only 13 papers address user bias, originating from human decisions during requirements gathering, design, and user interactions. The lower focus on user bias suggests it is either seen as less critical or more challenging to quantify and mitigate compared to model and data bias.

These numbers indicate that the research community prioritizes interventions within the model and data, one possible explanation is that practical solutions, like data augmentation, are more established. Addressing user bias may require broader cultural and organizational changes beyond technical interventions. As fairness research evolves, there may be opportunities to more comprehensively address user bias by integrating fairness considerations throughout the whole process.

\begin{table}
\setlength{\tabcolsep}{4.7pt}
\vspace{-3mm}
\centering
\caption{Bias type}
\label{tab:bias_type}
\begin{tabular}{+c^c^c}
\toprule
\rowstyle{\bfseries}
Bias Type & Amount of Studies & Studies \\
\midrule
User Bias & 13 & \makecell{S4, S17, S19, S28, S36, S41, S47, S66, S67, S68, S71, S79, S87} \\
\midrule
Data Bias & 34 & \makecell{S1, S3, S5, S6, S8, S9, S10, S21, S25, S26, S27, S28, S31, S36, S38, S40, S41, S42, S43, S44, S55, \\ S56, S57, S60, S62, S63, S75, S82, S85, S86, S87, S88, S93, S94} \\
\midrule
Model Bias & 64 & \makecell{S2, S5, S6, S7, S8, S9, S10, S11, S12, S13, S14, S15, S16, S18, S20, S22, S23, S24, S28, S29, S30, S32, \\ S33, S34, S35, S36, S37, S39, S41, S44, S45, S46, S48, S49, S50, S51, S52, S53, S54, S58, S59, S60, \\ S61, S64, S65, S69, S72, S73, S74, S75, S76, S77, S78, S80, S81, S83, S84, S87, S89, S90, S91, \\ S92, S94, S95} \\
\bottomrule 
\end{tabular}
\end{table}

\subsubsection{Fairness Metrics}

Fairness metrics are quantitative measures to evaluate whether a model behaves fairly.  Table~\ref{tab:fairness_metrics} highlights that metrics related to group fairness dominate the literature, comprising four of the five rows. In contrast, metrics addressing individual fairness lack a clear trend; although various studies mention individual fairness metrics, they are applied inconsistently, and no single metric stands out. Additionally, counterfactual fairness appears only in two studies. This disparity underscores a significant gap in developing consistent metrics for individual and counterfactual fairness. It also suggests a research focus on group fairness, potentially overlooking the importance of ensuring fairness at the individual level, consequently, the lack of consensus in evaluating individual fairness may undermine its reliability.


\begin{table}
\setlength{\tabcolsep}{4.7pt}
\vspace{-3mm}
\caption{Fairness metrics}
\label{tab:fairness_metrics}
\begin{tabular}{+c^c^c}
\toprule
\rowstyle{\bfseries}
Fairness Metrics & Amount of Studies & Studies \\
\midrule
Average Odds Difference (AOD) and related & 26 & \makecell{S1, S3, S4, S5, S6, S8, S9, S12, S14, S16, S24, S29, S36, S40, S41, \\ S43, S44, S46, S56, S70, S81, S87, S89, S90, S91, S93} \\
\midrule
\makecell{Statistical Parity Difference (SPD), \\ Demographic Parity (DP) and related} & 25 & \makecell{S1, S3, S4, S5, S6, S8, S9, S12, S16, S21, S24, S25, S36, S41, S43, \\ S44, S46, S56, S72, S81, S86, S87, S89, S90, S93} \\
\midrule
Equal Opportunity Difference (EOD) and related & 24 & \makecell{S1, S3, S4, S5, S6, S8, S9, S14, S16, S24, S29, S36, S40, S41, S43, \\ S70, S71, S81, S84, S87, S89, S90, S91, S93} \\
\midrule
Disparate Impact (DI) & 13 & \makecell{S1, S3, S6, S9, S21, S25, S41, S43, S46, S70, S81, S90, S93} \\
\midrule
Counterfactual Fairness & 2 & \makecell{S4, S41} \\
\bottomrule
\end{tabular}
\end{table}

\subsubsection{Targeted Approach}

Targeted Approach refers to strategies employed by the papers to address fairness-related issues, particularly in the context of bias. The approach target can be categorized into three main purposes:

\begin{itemize}
    
\item \textbf{Finding Bias}: in this case, the goal of the paper is to identify and detect biases within datasets, algorithms, models, or process. The papers focus on methods and techniques for uncovering potential sources of bias, whether in training data, model behavior, or decision-making processes.

\item \textbf{Mitigating Bias}: here, the purpose is to reduce or eliminate bias once it has been detected. The focus is on developing approaches or interventions that can either correct biased outcomes or prevent biased behaviors from emerging in the first place. These approaches might involve reweighting data, adjusting models, or applying fairness constraints.

\item \textbf{Explaining Bias}: this action aimes to understand the underlying causes and mechanisms behind the bias. The focus is on providing deeper insights into why bias occurs in particular models or datasets and developing a comprehensive understanding of the factors contributing to biased behavior. This step is crucial for developing long-term, sustainable strategies to address fairness in machine learning and systems.

\end{itemize}

These three approach target - finding, mitigating, and explaining bias — are critical to ensuring fairness in software systems. Together, they form a comprehensive approach to addressing biases and enhancing the ethical integrity and equitable outcomes of systems. The results are not exclusive; this means a paper might focus on finding bias while also mitigating and explaining it.

The Table~\ref{tab:action}  reveals a strong emphasis on Finding Bias (75 out of 95), followed by efforts in Mitigating Bias (56 out of 95), with fewer studies dedicated to Explaining Bias (21 out of 95). This pattern suggests that detecting and correcting biases are prioritized in fairness research, while understanding the underlying causes of bias is less developed. The papers approaching techniques to finding bias have become a core aspect of the field, with the major part of the studies associated, a beneficial result of this is dedicating efforts to finding biases allows practitioners to understand where fairness issues may lie, setting the stage for mitigation efforts. The lower representation in explaining bias suggests that while fairness research increasingly focuses on finding and fixing bias, understanding the “why” behind it is still an emerging area. Greater emphasis on explaining bias could offer valuable insights, enabling more targeted interventions and improved transparency.

\begin{table}
\setlength{\tabcolsep}{4.7pt}
\vspace{-3mm}
\caption{Targeted approach}
\label{tab:action}
\begin{tabular}{+c^c^c}
\toprule
\rowstyle{\bfseries}
Action & Amount of Studies & Studies \\
\midrule
Finding Bias & 75 & \makecell{S2, S3, S4, S5, S7, S8, S11, S12, S13, S14, S15, S16, S17, S18, S19, S20, S21, S22, S23, S24, S25, \\ S26, S27, S28, S30, S32, S33, S34, S35, S36, S37, S38, S39, S40, S41, S42, S43, S44, S45, \\ S46, S47, S49, S50, S51, S52, S55, S56, S58, S59, S60, S62, S65, S66, S69, S70, S71, S72, \\ S73, S75, S76, S77, S79, S80, S81, S82, S83, S85, S86, S87, S88, S90, S91, S93, S94, S95} \\
\midrule
Mitigating Bias & 56 & \makecell{S1, S3, S6, S9, S12, S14, S15, S16, S20, S21, S22, S23, S24, S26, S27, S28, S29, S31, S34, S35, \\ S36, S37, S38, S40, S41, S44, S45,  S46, S49, S50, S52, S54, S55, S58, S59, S62,  S69, S70, \\ S72, S73, S75, S76, S77, S79, S80, S81, S82, S83, S84, S86, S88, S89, S90, S91, S92, S93} \\
\midrule
Explaining Bias & 21 & \makecell{S3, S5, S17, S18, S28, S36, S41, S48, S52, S53, S57, S61, S63, S64, S65, S66, S67, \\ S68, S70, S74, S78} \\
\bottomrule
\end{tabular}
\end{table}

\subsection{Dataset and Model Characteristics}

This subgroup examines the two foundational components of machine learning systems: (i) dataset and (ii) model. We focus on identifying and cataloging the types of datasets and machine learning models commonly used in the studies. This includes examining which datasets are selected for training, testing, or analysis, and which machine learning models are employed across different applications. By understanding the commonalities and trends in dataset and model choices, we can gain insights into the typical assumptions, limitations, and research preferences that shape the field.

\subsubsection{Dataset}

The dataset play a crucial role in shaping the scope of a problem in machine learning and fairness research.
The most commonly used datasets across the literature are the Adult Census Income and German Credit, Table~\ref{tab:dataset}, which may suggest a focus on financial and socioeconomic problems in fairness research. This preference may reflects the nature and perspectives of many studies, as researchers often address fairness issues within these domains.
Different datasets can reveal varying aspects of fairness. For example, a dataset focused on hiring decisions might highlight gender or racial biases in employment, while a dataset related to healthcare outcomes could uncover disparities in medical treatment across different demographic groups. By exploring diverse datasets across different domains such as finance, education, and criminal justice, researchers and practitioners can gain new insights into fairness challenges and develop more robust fairness metrics and mitigation strategies.
Additionally, analyzing multiple datasets allows for a broader understanding of fairness across contexts. A fairness approach that works well in one domain may not necessarily translate effectively to another. Therefore, studying various datasets encourages the development of more generalizable fairness methodologies, helping ensure that fairness interventions are not just tailored to a single scenario but applicable across different real-world situations.

\begin{table}
\setlength{\tabcolsep}{4.7pt}
\vspace{-3mm}
\caption{Dataset}
\label{tab:dataset}
\begin{tabular}{+c^c^c}
\toprule
\rowstyle{\bfseries}
Dataset & Amount of Studies & Source \\
\midrule
Adult Census Income & 48 & Ref: \cite{adult_2} \\
\midrule
German Credit & 37 & Ref: \cite{statlog_(german_credit_data)_144} \\
\midrule
COMPAS / ProPublica & 30 & Ref: \cite{propublica_compas_2016} \\
\midrule
Bank Marketing & 27 & Ref: \cite{bank_marketing_222} \\
\midrule
Default of Credit Card Clients & 18 & Ref: \cite{default_of_credit_card_clients_350} \\
\midrule
Medical Expenditure Panel Survey & 17 & Ref: \cite{meps} \\
\midrule
Heart Health / Heart Disease & 14 & Ref: \cite{heart_disease_uci} \\
\midrule
Student Performance & 11 & Ref: \cite{student_performance_320} \\
\midrule
Diabetes & 5 & Ref: \cite{diabetes_34} \\
\bottomrule
\end{tabular}
\end{table}

\subsubsection{Model}

The model is a fundamental and critical component of machine learning systems and their fair behavior. Table~\ref{tab:model} shows a preference for Neural Networks (31 out of 95) and Logistic Regression (31 out of 95). The popularity of Logistic Regression may reflect a continued focus on interpretable models, whose transparency facilitates bias diagnosis, explanation, and mitigation. In contrast, the frequent use of Neural Networks suggests a growing interest in addressing fairness in complex, high-capacity models, which are often less interpretable but widely adopted in real-world applications. This dual emphasis highlights the field’s twofold direction: developing fairness-aware techniques that are both explainable for simpler models and scalable to deep architectures, ensuring that fairness principles can be effectively applied across diverse AI systems.

\begin{table}
\setlength{\tabcolsep}{4.7pt}
\vspace{-3mm}
\caption{Model}
\label{tab:model}
\begin{tabular}{+c^c^c}
\toprule
\rowstyle{\bfseries}
Model & Amount of Studies & Studies \\
\midrule
NN and DNN & 31 & \makecell{S2, S5, S7, S8, S9, S11, S15, S22, S26, S32, S33, S35, S36, S45, S46, S51, S52, S55, S56, \\ S58, S65, S69, S77, S83, S86, S88, S89, S90, S91, S94, S95} \\
\midrule
Logistic Regression & 31 & \makecell{S1, S3, S5, S6, S8, S10, S13, S14, S16, S19, S20, S21, S24, S27, S29, S30, S36, S40, S43, \\ S44, S51, S54, S65, S81, S83, S84, S86, S89, S90, S93, S94} \\
\midrule
Random Forest & 24 & \makecell{S1, S3, S5, S6, S8, S13, S16, S19, S22, S23, S25, S27, S29, S30, S33, S36, \\ S51, S61, S83, S89, S90, S93, S94, S95} \\
\midrule
Support Vector Machine & 20 & \makecell{S3, S6, S8, S16, S20, S21, S22, S27, S29, S33, S36, S44, S51, S65, S74, S86, \\ S88, S89, S90, S94} \\
\midrule
Decision Tree & 18 & \makecell{S13, S19, S20, S22, S24, S25, S29, S30, S33, S44, S51, S65, S83, S84, S86, S93, S94, S95} \\
\midrule
Naive Bayes & 10 & \makecell{S2, S5, S13, S19, S20, S21, S30, S51, S83, S86} \\
\bottomrule
\end{tabular}
\end{table}

\begin{tcolorbox}[colframe=black!50, colback=white, sharp corners, boxrule=0.5pt, width=\linewidth]
\textit{Main findings:}
\begin{itemize}
    \item Research focuses on post-processing and technical stages, while less attention is given to early-stage, less technical aspects, such as planning.
    \item Group fairness receives more attention, meaning issues related to groups are more likely to be addressed first.
    \item There is no clear pattern in individual fairness metrics; they are widely dispersed.
    \item Identifying bias receives more attention than mitigating or explaining it. Just few studies focus on explaining bias, indicating a gap in understanding bias root causes.
    \item The problem perspective is centered on the use of socioeconomic datasets.
    \item Traditional and less complex models are widely used.
\end{itemize}
\end{tcolorbox}

\section{Results -- Viability for Industrial Application}
\label{viability_industrial_application}

We analyzed the potential for industry adoption from four different perspectives: (1) Technology Readiness Level (TRL), (2) Industrial and Academic Engagement, (3) Tool Support, and (4) Availability to the Community.

\subsection{Technology Readiness Level (TRL)} 
TRL is a measure that indicates the maturity level of a technology or solution. It is represented by a number ranging from 1 to 9, with 1 being the least mature and 9 being the most mature. This metric was introduced by the Horizon 2020 European Commission \cite{mankins1995technology}.

\begin{figure}
    \centering
    \includegraphics[width=0.7\linewidth]{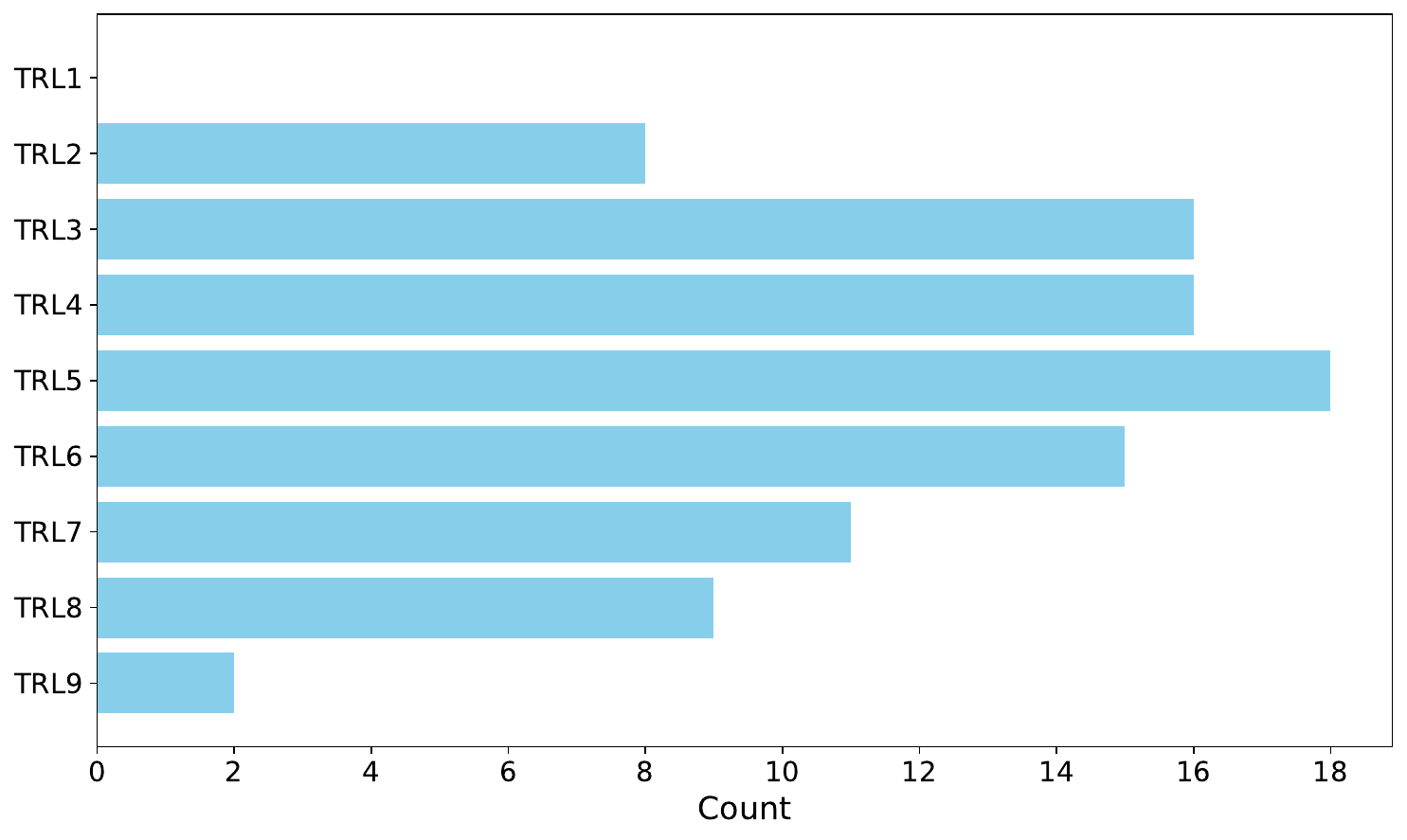}
    \caption{Distribution of TRL }
    \label{fig:trl}
\end{figure}

Figure~\ref{fig:trl} illustrates the distribution of studies across the TRL stages. The majority of studies (50 out of 95) are concentrated in stages 3, 4, and 5, indicating a low TRL, where technologies are primarily formulated, validated in a laboratory setting.
Additionally, 26 studies fall within stages 6 and 7, representing a medium TRL, where technologies are validated or demonstrated in a relevant environment.
Finally, 11 studies have a high TRL, distributed across stages 8, and 9, where technologies are either fully developed, demonstrated, or proven in an operational environment.

Based on these results, we can conclude that: most research on fairness solutions is still in its early stages concerning the transferability of developed technologies to industry; a significant number of studies (34 out of 95) focus on proof of concept (TRL 3) or have validated the technology concept (TRL 5); only a few studies have demonstrated the technology in an operational environment; none of the studies are solely focused on discussing the basic principles of fairness.

\subsection{Industrial and Academic Engagement} In this category, we classify each study based on the affiliations of its authors. A study is considered academic if all authors are affiliated with universities or research centers, industrial if all authors are affiliated with companies, or both if the authors represent a combination of academia and industry.

\begin{figure}
    \centering
    \includegraphics[width=0.8\linewidth]{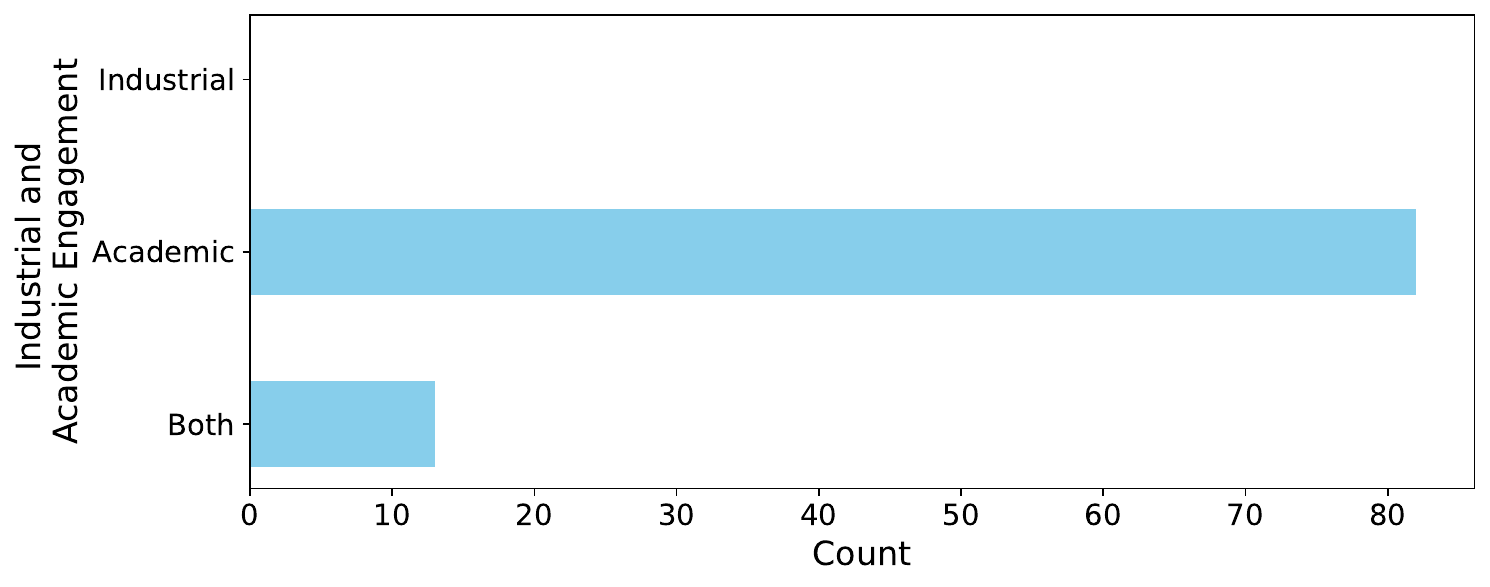}
    \caption{Distribution of ``industrial and academic engagement"}
    \label{fig:ii_involvement}
\end{figure}

In Figure \ref{fig:ii_involvement}, we observe that only 13 studies involved collaboration between academia and industry, while the majority (82 out of 95) were purely academic. This highlights a significant gap in knowledge exchange between these two sectors. Effective collaboration is crucial for addressing industrially relevant challenges, ensuring that research is aligned with real-world needs, and facilitating the transfer of technologies from academia to industry. The current landscape suggests that there is still considerable progress to be made in bridging this gap and fostering stronger industry-academia partnerships.

\subsection{Tool Support} 
\begin{figure}
    \centering
    \includegraphics[width=0.8\linewidth]{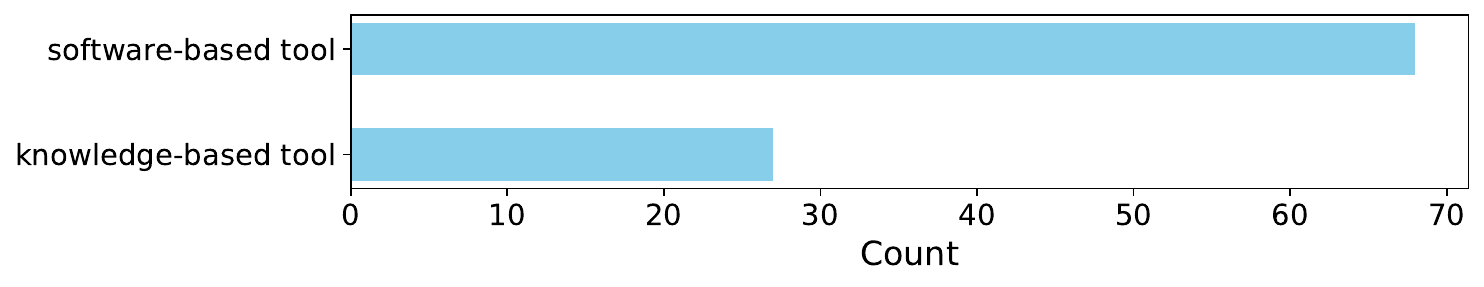}
    \caption{Distribution of ``tool support"}
    \label{fig:tool_support}
\end{figure}

This perspective assesses whether the solution is software-based or knowledge-based, following the approach of \cite{di2017research}. In Figure \ref{fig:tool_support}, we observe that the majority of the studies (68 out of 95) focus on software-based tools, while 27 out of 95 are knowledge-based. This indicates that fairness research has largely favored practical solutions to address bias-related challenges.

\subsection{Availability to the Community} This criterion assesses whether the system is openly accessible to the broader community, which can influence its adoption and collaborative development. We have a balanced distribution between studies available for community replication (55 out of 95) and those that are not openly accessible (40 out of 95).

\begin{figure}
    \centering
    \includegraphics[width=0.8\linewidth]{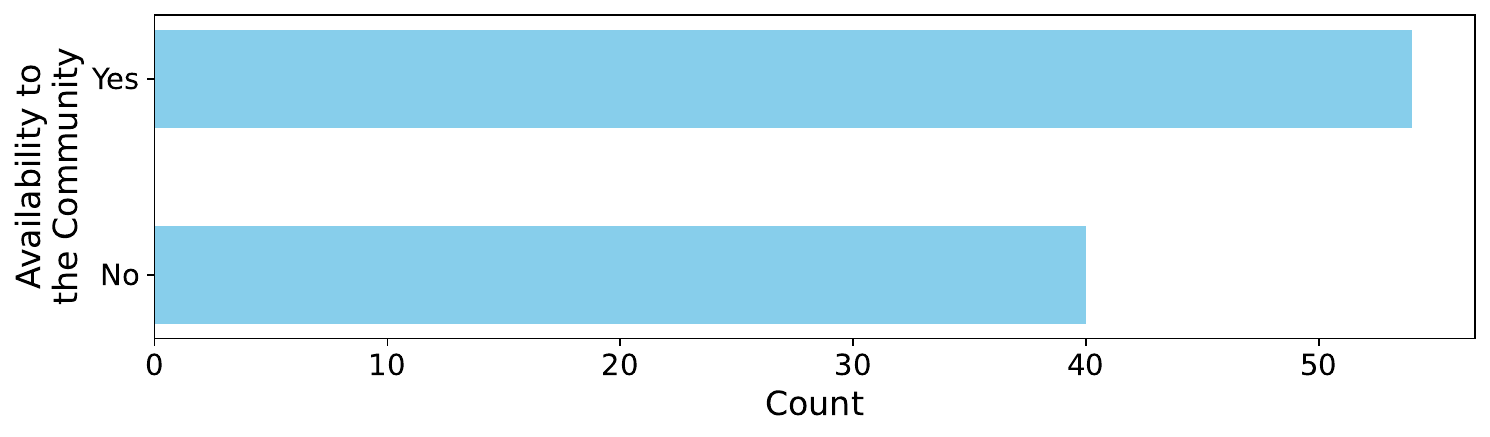}
    \caption{Distribution of ``availability to the community"}
    \label{fig:open}
\end{figure}

\begin{tcolorbox}[colframe=black!50, colback=white, sharp corners, boxrule=0.5pt, width=\linewidth]
\textit{Main findings:}
\begin{itemize}
    \item  The TRL scores of most studies indicate that industrial transferability remains distant.
    \item There is a significant gap in industry-academia collaboration, with most studies being purely academic, highlighting the need for better knowledge exchange and technology transfer.
\end{itemize}
\end{tcolorbox}

\section{Threats to Validity}
\label{threats_validity}

This section discuss potential limitations in our study that might impact the findings. It is essential for transparency, as it shows the awareness of any factors that could have influenced our results in unintended ways. It typically covers four types of validity: internal, external, construct, and conclusion.

Internal validity concerns factors within the study that might have influenced the results. It can be summarized by asking: How well was the study conducted internally?
In our study, we ensured internal validity by basing our methodology on the approach implemented in \cite{di2017research}. We used a robust and reliable data source (Scopus) which includes only peer-reviewed papers. We applied inclusion and exclusion criteria iteratively and incrementally to refine our selection of studies, and we also implemented automated snowballing technique to locate relevant studies that might not have appeared in the initial search results. These methods helped us meet the criteria for internal validity, minimizing bias and ensuring that our findings were directly attributable to our research approach.

External validity relates to the generalizability of the findings to other settings, populations, or times. The core question here is: How well can the study results apply outside of this specific study?
In the Study Design section, we provided a step-by-step description of our methodology to ensure its replicability. Additionally, we have made all scripts and artifacts also available, allowing others to reproduce our method. Based on the methodology used here and grounded in \cite{di2017research}, we believe our approach has the necessary generalizability to be repeated and applied across different contexts and research areas. This increases the study’s relevance beyond the current dataset and setting.

Construct validity checks if the study actually measures what it claims to measure.
In our study, we aimed to answer the following research questions:
\textit{a) What are the leading solutions proposed by researchers to achieve fairness in software systems?; 
b) What are the primary research focuses within this field?;
c) How are these solutions integrated within the context of Software Engineering?
d)  What is the viability of the industrial context? }
Through detailed analysis and discussion of the data tables, we can ensure that our results directly addressed these initial questions. By evaluating the outcomes in relation to these specific research goals, we can say that our study effectively measures the constructs it intended to explore.

Conclusion validity is about whether the conclusions drawn from the study data are solid and accurate. It considers whether researchers made correct inferences based on the data and whether any statistical errors might have affected the conclusions.
In our study, our conclusions were drawn from a careful analysis and logical discussion of the data tables. We examined the patterns and relationships in the data systematically to ensure our interpretations were well-founded. This approach minimized the risk of misinterpretation and reinforced the reliability of our conclusions.

\section{Related Works}
\label{related_works}

Interest in achieving fairness in systems has grown significantly. Numerous studies focus on understanding, measuring, and mitigating biases in the applications. This section contextualizes the contributions of our systematic mapping study by comparing it with previous studies.

Balayn et al. \cite{balayn2021managing} emphasize that good data management is key to fairness. They review methods to handle bias in systems that make decisions based on data. Unlike our approach, which organizes solutions in a more broader way and also according to different stages of software development, they focus on the data itself. They argue that fairness should be built into how we handle and manage data, not just applied to the outputs of machine learning models.

Chen et al. \cite{chen2024fairness} provide a detailed survey on testing for fairness in machine learning, on the other hand, our paper maps fairness research in software engineering, considering aspects like research focus and trends. Unlike [26], which focuses on fairness testing as a non-functional requirement, using a benchmarking-like method to compare bias mitigation techniques, our study introduces a framework to holistically analyze Software Fairness papers, offering a more comprehensive perspective across all stages of software development.

Serban et al. \cite{serban2005journal} look into how engineering practices are adopted in machine learning, especially trustworthy AI practices like fairness. They focus on how these practices are used during development and deployment, providing insights into team behaviors and how effective these practices seem. Our study goes further by creating a structured framework that covers various solution approaches, including algorithmic methods and ways to reduce bias, helping to analyze research trends in fairness solutions more systematically.

Soremekun et al. \cite{soremekun2022software} study fairness as a feature of software, creating a classification based on fairness measures, tasks, and levels of access to the system (like black-box or white-box analysis). While their study provides insights into fairness issues in specific software analysis methods, our framework looks at fairness solutions at different stages of the workflow, characteristics of the dataset, and model considerations. This places fairness within a complete software engineering perspective that goes beyond specific fairness analysis techniques.

Pessach et al. \cite{pessach2022review} focus on fairness in machine learning, explaining what fairness means and the balance between accuracy and fairness. They categorize fairness methods into pre-processing, in-processing, and post-processing, similar to our way of organizing solutions. However, our study also includes, an industrial analysis, and software-specific fairness aspects like testing and verification, offering a framework that aligns with the stages of software engineering to help future real-world applications.

Ntoutsi et al. \cite{ntoutsi2020bias} present a broad survey on bias in AI systems that use data, grouping methods into understanding bias, mitigating it, and accounting for it. Their multidisciplinary approach emphasizes the ethical implications of AI fairness but doesn't specifically focus on software engineering aspects. In contrast, our study provides a structured view tailored to software engineering applications, offering practical insights into implementing fairness solutions within software systems.

Sesari et al. \cite{sesari2024understanding} explore discussions about fairness on Stack Exchange to understand software professionals' concerns within a social and technical context. They find that software engineers often discuss fairness related to income, hiring, and workplace treatment. By focusing on social and human aspects of fairness, they use a qualitative, discussion-based approach. This differs from our structured classification of fairness methods across empirical studies, algorithms, and frameworks. While their insights provide valuable context on fairness as a workplace issue, our study aims to guide technical implementations in software systems that are aware of fairness.

By creating a classification framework and systematically organizing fairness research in software engineering, and also analyzing the industrial aspect, our study aims to offer a complete view of fairness solutions in the systems context. We connect theoretical concepts of fairness with their practical application in software development. This comparison highlights our contribution in placing fairness as an integral part of the software engineering process, guiding future research and real-world implementations in fairness-aware software systems.

\section{Conclusion}
\label{conclusion}

This study aimed to deliver a comprehensive analysis exploring the connections among various research efforts addressing fairness solutions in software. Through a systematic review of 95 studies, we offered an insightful overview of the current landscape in software fairness research. The findings of this study serve as a resource for researchers seeking to advance contributions in this domain and for practitioners looking to gain a deeper understanding of established research. We structured our final thoughts on the conclusion based on the research inquiries outlined in the Study Design section.

\begin{enumerate}[label=\alph*)]
    \item \textit{What are the current leading solutions proposed by researchers to achieve fairness in AI systems?}

    In this work, using the framework described in Study Design, we identified 95 studies addressing fairness in the context of systems, that can be found in \cite{nepomuceno_2025_17435899}. By analyzing the data tables presented in the Results section, we drew several key insights regarding the research perspectives on this topic:
    \begin{itemize}
    \item Researchers have focused on developing frameworks, methods, and algorithms to address fairness.
    \item The results indicate that most proposed solutions are applied during the post-processing stage of development. One possible reason is that this stage offers greater flexibility and simplification when verifying fairness requirements. However, it is equally important to ensure that fairness considerations extend to other stages of the data processing pipeline, such as pre-processing.
    \item Neural networks and traditional models such as logistic regression and random forests are frequently used to diagnose bias. The widespread preference for traditional models can be attributed to their interpretability and effectiveness in fairness evaluations, underscoring their central role in current fairness research.
    \end{itemize}
    
    \item \textit{What are the primary research focuses within this field?}

    One of the steps in the Framework (Section 2), is the keywording technique, which retrieves the core topics and concepts of a domain. Using this approach, we identified the principal topics currently discussed in software fairness. We defined them as items, clustered, and analyzed them in table format in the Results section. The identified items are shown highlighted in Figure \ref{fig:researchfocus}. Key findings from this analysis include:
    
    \begin{itemize}
        \item (Focus on Group Fairness Metrics) An area needing further exploration is the full spectrum of fairness metrics. According to the results, the literature demonstrates a strong emphasis on group fairness metrics, leaving individual fairness and counterfactual fairness underexplored. These metrics hold the potential to provide more personalized and context-aware fairness solutions, addressing nuanced biases that group metrics may overlook. Expanding research into these areas can enrich the diversity of fairness approaches and broaden their applicability.
        \item (Use of Common Datasets) A limitation lies in the restricted domain diversity of existing studies. The frequent reliance on financial and socioeconomic datasets (Adult Census Income and German Credit) indicates a narrow scope, while fairness challenges in critical areas such as healthcare and education remain less addressed. Future work must prioritize domain-specific studies to uncover unique fairness issues and tailor solutions to diverse real-world.
        \item (Focus on Bias Detection) The research also highlights an insufficient focus on explaining the root causes of biases. While many studies excel in finding and removing bias, understanding its origins remains an underdeveloped area. Developing explanation tools can provide deeper insight into why biases occur, enabling more targeted and effective mitigation strategies.
        \item (Focus on Model and Data Bias) Researchers are mostly addressing biases originating from models and data, with less attention given to “user bias”.  This highlights a potential gap in the literature, as user bias introduced during the early stages of design and development can significantly influence fairness outcomes down the line. This indicates a potential gap in the research, as “user bias”, introduced during the early design and development stages, can have significant downstream effects on the fairness outcomes.
        \item (Focus on Later Stages of Development) One of the key findings is the uneven distribution of fairness efforts across the software development life-cycle. Current research predominantly focuses on technical stages, with limited attention to early stages. This reactive approach to fairness often leads to addressing biases after they have manifested, rather than preventing them at the source. A shift toward embedding fairness considerations in early-stage activities, such as requirement engineering and system architecture design, is essential for proactive and sustainable solutions.
    \end{itemize}

    \item \textit{How do these solutions integrate within the context of Software Engineering?}

    In the context of software engineering, solutions are primarily integrated during the Implementation and Testing phases of the software development life-cycle. These stages likely receive the most attention due to their technical complexity and the wide range of tasks they involve, making them a natural starting point for researchers addressing fairness issues. However, there is limited integration of fairness considerations during early stages, like the Planning phase, suggesting that the early stages of software development are underrepresented in current research. The Deployment and Maintenance stages also receive minimal attention, which is critical since fairness issues can persist or emerge post-deployment. The overall integration indicates that while there is a general concern for addressing fairness across different stages, there is a need for more comprehensive inclusion of fairness principles throughout the entire software development.
    
    \item \textit{What is the viability of the industrial context?}

    The feasibility of fairness solutions in an industrial context remains limited but promising based on the following factors:
    \begin{itemize}
        \item Technology readiness level (TRL) is mostly low to medium -- This suggests that while fairness solutions are being developed, they are not yet widely deployable in industry.
        \item Weak industry-academia collaboration -- Limited collaboration slows the transfer of research into practical industrial applications, reducing feasibility.
        \item Strong emphasis on software-based solutions -- This increases the feasibility of adoption since industries favor concrete, implementable tools.
        \item Mixed open-source availability -- About half of the studies provide open access, allowing for potential adaptation and integration into industry. The other half remain closed, limiting accessibility and slowing adoption.
    \end{itemize}

    The viability of fairness solutions in the industrial context is currently low to moderate due to the early maturity of technologies and limited industry collaboration. However, the presence of software-based tools and increasing open accessibility indicate that, with further development and stronger partnerships, adoption potential can improve significantly.
    
\end{enumerate}

To address the gaps mentioned, the field needs to embrace a holistic approach that integrates fairness throughout the software development life-cycle. This includes early-stage interventions, comprehensive fairness metrics, causal explanation frameworks, domain-specific research, and tools for real-world deployment and sustained monitoring. Collaborations with industry partners can further bridge the gap between theoretical research and practical application.

In conclusion, while significant progress has been made in understanding and addressing fairness in software systems, the path forward requires actionable studies and innovative approaches. By addressing the identified gaps, researchers and practitioners can develop inclusive, transparent, and equitable AI systems that align with societal values and ethical patterns.

\newpage
\appendix
\section{Research studies analyzed in the systematic literature mapping}
\label{app1}

\small
\begin{longtable}{@{}c c c p{10cm}@{}}
\caption{Studies selected for the systematic literature mapping} \\
\toprule
\textbf{ID} & \textbf{Year} & \textbf{Venue} & \textbf{Title} \\
\midrule
\endfirsthead

\toprule
\textbf{ID} & \textbf{Year} & \textbf{Venue} & \textbf{Title} \\
\midrule
\endhead

\midrule
\multicolumn{4}{r}{\textit{Continued on next page}} \\
\endfoot

\bottomrule
\endlastfoot

S1 & 2021 & Conference & Ignorance and Prejudice in software fairness \\
S2 & 2022 & Journal & Astraea: Grammar-Based Fairness Testing \\
S3 & 2021 & Conference & Bias in machine learning software: Why? how? what to do? \\
S4 & 2023 & Journal & Fairness in Design: A Framework for Facilitating Ethical Artificial Intelligence Designs \\
S5 & 2022 & Journal & FairMask: Better Fairness via Model-Based Rebalancing of Protected Attributes \\
S6 & 2023 & Conference & Fix Fairness, Don't Ruin Accuracy: Performance Aware Fairness Repair using AutoML \\
S7 & 2023 & Conference & Information-Theoretic Testing and Debugging of Fairness Defects in Deep Neural Networks \\
S8 & 2024 & Conference & Fairness Improvement with Multiple Protected Attributes: How Far Are We? \\
S9 & 2024 & Journal & An Empirical Study on Correlations between Deep Neural Network Fairness and Neuron Coverage Criteria \\
S10 & 2023 & Journal & Fair Enough: Searching for Sufficient Measures of Fairness \\
S11 & 2020 & Conference & Perfectly parallel fairness certification of neural networks \\
S12 & 2024 & Conference & RUNNER: Responsible UNfair NEuron Repair for Enhancing Deep Neural Network Fairness \\
S13 & 2020 & Journal & Automatic Fairness Testing of Machine Learning Models \\
S14 & 2020 & Conference & Fairway: A way to build fair ML software \\
S15 & 2023 & Conference & Fairify: Fairness Verification of Neural Networks \\
S16 & 2022 & Conference & MAAT: a novel ensemble approach to addressing fairness and performance bugs for machine learning software \\
S17 & 2024 & Journal & MBFair: a model-based verification methodology for detecting violations of individual fairness \\
S18 & 2020 & Conference & Metamorphic testing and certified mitigation of fairness violations in NLP models \\
S19 & 2024 & Conference & ReFAIR: Toward a Context-Aware Recommender for Fairness Requirements Engineering \\
S20 & 2017 & Conference & Fairness testing: Testing software for discrimination \\
S21 & 2022 & Conference & Training Data Debugging for the Fairness of Machine Learning Software \\
S22 & 2018 & Conference & Automated directed fairness testing \\
S23 & 2022 & Journal & Search-based fairness testing for regression-based machine learning systems \\
S24 & 2024 & Journal & Search-based Automatic Repair for Fairness and Accuracy in Decision-making Software \\
S25 & 2019 & Conference & The Impact of Data Preparation on the Fairness of Software Systems \\
S26 & 2020 & Conference & Improving Fairness in Speaker Recognition \\
S27 & 2022 & Conference & Identifying Imbalance Thresholds in Input Data to Achieve Desired Levels of Algorithmic Fairness \\
S28 & 2023 & Journal & A Maturity Model for Trustworthy AI Software Development \\
S29 & 2022 & Conference & Fairness-aware Configuration of Machine Learning Libraries \\
S30 & 2024 & Journal & Diversity-aware fairness testing of machine learning classifiers through hashing-based sampling \\
S31 & 2023 & Conference & Exploring and Repairing Gender Fairness Violations in Word Embedding-based Sentiment Analysis Model through Adversarial Patches \\
S32 & 2021 & Conference & Efficient white-box fairness testing through gradient search \\
S33 & 2021 & Journal & Coverage-Guided Fairness Testing \\
S34 & 2021 & Conference & Individual Fairness for Graph Neural Networks: A Ranking based Approach \\
S35 & 2024 & Conference & MAFT: Efficient Model-Agnostic Fairness Testing for Deep Neural Networks via Zero-Order Gradient Search \\
S36 & 2023 & Journal & A Comprehensive Empirical Study of Bias Mitigation Methods for Machine Learning Classifiers \\
S37 & 2018 & Journal & Supporting Many-Objective Software Requirements Decision: An Exploratory Study on the Next Release Problem \\
S38 & 2023 & Journal & Multi-objective search for gender-fair and semantically correct word embeddings \\
S39 & 2024 & Journal & Software doping analysis for human oversight \\
S40 & 2024 & Journal & FairBalance: How to Achieve Equalized Odds With Data Pre-processing \\
S41 & 2021 & Conference & The landscape and gaps in open source fairness toolkits \\
S42 & 2023 & Journal & Measuring Imbalance on Intersectional Protected Attributes and on Target Variable to Forecast Unfair Classifications \\
S43 & 2022 & Workshop & Fair-SSL: Building fair ML Software with less data \\ 
S44 & 2021 & Conference & Fairea: A model behaviour mutation approach to benchmarking bias mitigation methods \\
S45 & 2020 & Conference & White-box fairness testing through adversarial sampling \\
S46 & 2021 & Journal & Are My Deep Learning Systems Fair? An Empirical Study of Fixed-Seed Training \\
S47 & 2020 & Journal & Algorithmic bias and the value sensitive design approach \\
S48 & 2023 & Workshop & FairnessLab: A Consequence-Sensitive Bias Audit and Mitigation Toolkit \\
S49 & 2023 & Journal & Balanced Fair K-Means Clustering \\
S50 & 2021 & Journal & Fair near neighbor search via sampling \\
S51 & 2019 & Conference & Testing machine learning algorithms for balanced data usage \\
S52 & 2020 & Journal & A responsible machine learning workflow with focus on interpretable models, post-hoc explanation, and discrimination testing \\
S53 & 2023 & Journal & Bias in human data: A feedback from social sciences \\
S54 & 2019 & Conference & Fairness without harm: Decoupled classifiers with preference guarantees \\
S55 & 2021 & Conference & AI Extension of SQuaRE Data Quality Model \\
S56 & 2023 & Journal & Data collection and quality challenges in deep learning: a data-centric AI perspective \\
S57 & 2021 & Conference & Detecting Discrimination Risk in Automated Decision-Making Systems with Balance Measures on Input Data \\
S58 & 2023 & Journal & QuoTe: Quality-oriented Testing for Deep Learning Systems \\
S59 & 2022 & Conference & An Algorithmic Framework for Bias Bounties \\
S60 & 2023 & Journal & Tiny, Always-on, and Fragile: Bias Propagation through Design Choices in On-device Machine Learning Workflows \\
S61 & 2020 & Conference & Experimental study on generating multi-modal explanations of black-box classifiers in terms of gray-box classifiers \\
S62 & 2023 & Conference & A Multidimensional Analysis of Social Biases in Vision Transformers \\
S63 & 2022 & Conference & De-biasing "bias" measurement \\
S64 & 2023 & Journal & Monitoring Algorithmic Fairness Under Partial Observations \\
S65 & 2021 & Journal & Evaluating impact of race in facial recognition across machine learning and deep learning algorithms \\
S66 & 2023 & Journal & Integrating Fairness in the Software Design Process: An Interview Study With HCI and ML Experts \\
S67 & 2018 & Conference & A qualitative exploration of perceptions of algorithmic fairness \\
S68 & 2022 & Conference & Assessing the Fairness of AI Systems: AI Practitioners' Processes, Challenges, and Needs for Support \\
S69 & 2024 & Journal & Causal Intervention for Fairness in Multibehavior Recommendation \\
S70 & 2023 & Journal & A step toward building a unified framework for managing AI bias \\
S71 & 2021 & Journal & Software documentation is not enough! Requirements for the documentation of AI \\
S72 & 2023 & Conference & RELIANT: Fair Knowledge Distillation for Graph Neural Networks \\
S73 & 2020 & Journal & Certified Machine-Learning Models \\
S74 & 2023 & Journal & Abstract Interpretation-Based Feature Importance for Support Vector Machines \\
S75 & 2021 & Conference & Slice Tuner: A Selective Data Acquisition Framework for Accurate and Fair Machine Learning Models \\
S76 & 2019 & Conference & Proportionally fair clustering \\
S77 & 2019 & Journal & The what-if tool: Interactive probing of machine learning models \\
S78 & 2022 & Journal & Pros and cons of GAN evaluation measures: New developments \\
S79 & 2021 & Conference & Non-functional Requirements for Machine Learning: Understanding Current Use and Challenges in Industry \\
S80 & 2024 & Workshop & CO*IR: A Greedy and Individually Fair Re-ranker \\
S81 & 2022 & Conference & Towards model-based bias mitigation in machine learning \\
S82 & 2021 & Journal & Imbalanced data as risk factor of discriminating automated decisions: A measurement-based approach \\
S83 & 2021 & Conference & MLCHECK-property-driven testing of machine learning classifiers \\
S84 & 2025 & Journal & Multi-objective swarm intelligence approach for bias mitigation in decision-making software \\
S85 & 2025 & Journal & Experience: Bridging Data Measurement and Ethical Challenges with Extended Data Briefs \\
S86 & 2025 & Journal & FairPlay: A Collaborative Approach to Mitigate Bias in Datasets for Improved AI Fairness \\
S87 & 2025 & Journal & How fair are we? From conceptualization to automated assessment of fairness definitions \\
S88 & 2024 & Workshop & Method for analysis and formation of representative text datasets \\
S89 & 2025 & Conference & Diversity Drives Fairness: Ensemble of Higher Order Mutants for Intersectional Fairness of Machine Learning Software \\
S90 & 2025 & Conference & Fairness Testing Through Extreme Value Theory \\
S91 & 2025 & Journal & Fairness Mediator: Neutralize Stereotype Associations to Mitigate Bias in Large Language Models \\
S92 & 2025 & Conference & Building Bridges, Not Walls: Fairness-Aware and Accurate Recommendation of Code Reviewers via LLm-Based Agents Collaboration \\
S93 & 2025 & Journal & FairPreprocessor: Better Fairness via Addressing Imbalanced Data through Synthetic Data Generation and Mitigating Biased Labels \\
S94 & 2024 & Journal & Beyond the Seeds: Fairness Testing via Counterfactual Analysis of Non-Seed Instances \\
S95 & 2024 & Conference & RSUTT: Robust Search Using T-Way Testing \\
\end{longtable}

\bibliographystyle{elsarticle-num} 
\bibliography{references}

\end{document}